\documentclass[aps,prd,amsmath,amssymb,preprintnumbers,twocolumn]{revtex4}
\usepackage[a4paper,left=20mm,right=20mm,top=25mm,bottom=25mm]{geometry}
\usepackage{amsmath}
\usepackage{amssymb}
\usepackage[pdftex]{color,graphicx}
\usepackage{slashed}
\usepackage{feynmf}
\usepackage{rotating}
\usepackage{graphicx}
\usepackage{booktabs}

\newcommand{\Tr}{\text{Tr}}

\newcommand{\SO}{\text{SO}}
\newcommand{\SU}{\text{SU}}
\newcommand{\Sp}{\text{Sp}}
\newcommand{\U}{\text{U}}

\newcommand{\refeq}[1]{(\ref{#1})}

\usepackage[colorlinks,linkcolor=blue,citecolor=blue,urlcolor=blue,linktocpage]{hyperref}
\newcommand{\hhref}[2][]{\href{http://arxiv.org/abs/#2#1}{arXiv:#2}}

\allowdisplaybreaks[2]

\makeatletter
\def\l@subsection#1#2{}
\def\l@subsubsection#1#2{}
\makeatother

\begin{document}
\title{\LARGE WIMP Dark Matter in Composite Higgs Models and the Dilaton Portal}
\author{Manki Kim}
\email{visionk@kaist.ac.kr}
\affiliation{Department of Physics, Korea Advanced Institute of Science and Technology, 335 Gwahak-ro, Yuseong-gu, Daejeon 305-701, Korea}
\author{Seung J. Lee}
\email{sjjlee@korea.edu}
\affiliation{Department of Physics, Korea University, Seoul 136-713, Korea}
\affiliation{School of Physics, Korea Institute for Advanced Study, Seoul 130-722, Korea}
\author{Alberto Parolini}
\email{parolini85@kias.re.kr} 
\affiliation{$^d$Quantum Universe Center, Korea Institute for Advanced Study, Seoul 130-722, Korea}

\begin{abstract}
We study under which conditions a scalar particle is a viable WIMP Dark Matter candidate with Higgs and dilaton interactions. The theory is a composite Higgs model with top partial compositeness where both the Higgs and the Dark Matter candidate arise as pseudo Goldstone boson of the coset SO$(6)$/SO$(5)$ from a new physics sector. We highlight the role of the dilaton in direct and indirect searches. We find that a Dark Matter particle with a mass around 200-400 GeV and a relatively light dilaton are a fair prediction of the model. \\[2mm]
\end{abstract}

\maketitle

\section{Introduction}
Light scalars are believed to be unlikely in Nature, unless there is a fine tuning or there exists an underlying dynamics screening the quadratic ultraviolet sensitivity. Indeed the Standard Model (SM) suffers from the hierarchy problem because of the Higgs boson: an interesting possibility is that the Higgs boson, rather than an elementary particle, is a composite object, a bound state of a new, yet undiscovered, interacting theory which gets strong at the TeV scale. In particular the idea that the Higgs is not only a composite object but a pseudo Nambu Goldstone boson (pNGB), like pions in QCD, is especially appealing, because of the approximate built in shift symmetry.

From a different perspective, also the Dark Matter (DM) density in the Universe could be accounted for by a scalar particle, again subject to the same naturalness issue, and if it is a weakly interacting massive particle (WIMP), its mass should be broadly in the TeV range. Therefore a very compelling picture emerges if a single new strongly interacting sector is responsible for both the Higgs and the DM. We pursue this approach in a next to minimal pNGB Composite Higgs Model (CHM), based on the symmetry breaking coset $\SO(6)/\SO(5)$: it includes a custodial $\SO(4)$ and it is exactly described by five Goldstone modes, a bidoublet $H$ and a singlet $\eta$. This coset, or the isomorphic $\SU(4)/\Sp(4)$, can be formulated in an underlying theory of fundamental techni-quarks and it has already received some attention \cite{Katz:2005au,Gripaios:2009pe,Galloway:2010bp,Barnard:2013zea,Cacciapaglia:2014uja,Cacciapaglia:2015eqa}. If $\eta$ is sufficiently stable it is a perfect DM candidate: this is achieved if the theory respects a global $\mathbb{Z}_2$ symmetry under which $\eta$ is odd. The main difference with the case of elementary scalars is in the form of the interactions. This very predictive setup has already been explored \cite{Frigerio:2012uc,Marzocca:2014msa}. We want to extend the analysis assuming that the strong sector provides a second DM portal to SM particles: on top of Higgs exchange the dilaton could play an important role, if the strong sector is an approximate Conformal Field Theory (CFT) and it features a light dilaton. A light dilaton is also a rare phenomenon in spontaneously broken CFTs in the sense that it requires fine tuning, \cite{Bellazzini:2012vz,Chacko:2012sy,Chacko:2013dra,Bellazzini:2013fga,Coradeschi:2013gda}, but if present it affects the DM phenomenology, if it is a different state than the Higgs scalar. We will show how in our model the light dilaton affects the DM phenomenology, mainly fixing a lighter DM mass; moreover it gives the dominant contribution to Sommerfeld enhanced processes. The dilaton portal in composite DM models has been studied in \cite{Blum:2014jca}, but neglecting Higgs effects. A complete picture including both is the main object of our present work. In \cite{Efrati:2014aea} a similar interplay was studied, but without the pNGB structure.

The rest of the paper is organized as follows. After defining an effective Lagrangian in section \ref{sec: goldstone lagrangians}, including the other composite resonances typically considered in CHM, we introduce the dilaton field $\sigma$ and we derive the interactions between the light scalars, $h$, $\eta$ and $\sigma$, and the SM fermions and vectors in section \ref{sec:dilaton extension}. We move to DM properties, starting from the computation of the relic density, section \ref{sec: relic abundance}, to direct and indirect constraints, in section \ref{sec: direct detection} and \ref{sec: indirect detection} respectively. We take into account collider constraints in section \ref{sec: collider}. Finally we summarize and we draw our conclusions in section \ref{sec: summary}.  

\section{The $\bf\SO(6)/\SO(5)$ model}\label{sec: goldstone lagrangians}

\subsection{Scalar Sector}
The new physics sector, behaving as a CFT, is perturbed by a deformation, which becomes strong at an energy scale around the TeV. It possesses, in isolation, an approximate global $\SO(6)$ symmetry, spontaneously broken to $\SO(5)$. As a result five pseudo Goldstone bosons arise, a complex doublet $H$ and a singlet $\eta$. $H$ transforms as a bi-doublet under the custodial $\SU(2)_{L}\times\SU(2)_{R}\subseteq\SO(5)$, and $\eta$ is a singlet. According to the Callan Coleman Wess Zumino (CCWZ) formalism \cite{Coleman:1969sm,Callan:1969sn} the Lagrangian for the Goldstone bosons, in the unitary gauge, is written as
\begin{align}
\mathcal{L} =& \frac{f^2}{4} \Tr[d_{\mu}d^{\mu}]=\frac{f^2}{2}(D_{\mu}\Sigma)^{T}D^{\mu}\Sigma \\ 
=& \frac{1}{2}\left[\partial_{\mu}h \partial^{\mu}h+\partial_{\mu}\eta \partial^{\mu}\eta+\frac{(h\partial_{\mu}h+\eta \partial_{\mu} \eta)^2}{f^2-h^2-\eta^2}\right] \nonumber\\ &+ \frac{h^2}{8}(g_{0}^2((W^{1}_{\mu})^2+(W^{2}_{\mu})^2)+(g_{0}'B_{\mu}-g_{0}W^{3}_{\mu})^2 )\,.\nonumber
\end{align}
where $U(x)$ and$\ \Sigma$\ are defined in terms of the broken $\SO(6)$ generators $T^{\hat{a}}$ as
\begin{align}
U(x)=&e^{(i \sqrt{2}{\theta^{\hat{a}}}T^{\hat{a}})}\,,\,\,
\Sigma(x)= U(x) \left(0~0~0~0~0~1\right)^{T}\,,
\end{align}
and in the unitary gauge
\begin{equation}
\Sigma=\frac{1}{f}\left(0,0,h,0,\eta,\sqrt{f^2 -h^2 -\eta^2}\right)^{T}\,.
\end{equation}
$d_\mu^{\hat{a}}$ is defined as $i\Tr(U^\dagger\partial_\mu U T^{\hat{a}})$.
The scalar potential is radiatively generated once $\SO(6)$ breaking effects are included, namely once the strong sector is coupled to the SM, and it depends on the details of the composite sector and of the mixings, therefore it is model dependent. Nonetheless it can be parametrized in the following way:
\begin{align}\label{eq: Veff fermions}
V_{f}(h,\eta,\chi)=&\frac{\mu^{2}_{h,f}}{2}h^2  + \frac{\mu^{2}_{\eta,f}}{2}\eta^2 +\frac{\lambda_{h,f}}{4}h^4 +\frac{\lambda_{\eta,f}}{4}\eta^4 \nonumber\\
&+\frac{\lambda_{h\eta,f}}{4}\eta^2 h^2\,.
\end{align}
We limit to models in whose vacuum the ElectroWeak (EW) symmetry is broken 
\begin{equation}
h=\langle h \rangle+\sqrt{1-\xi}h_{phys}\,,\quad\langle\eta\rangle=0\,,
\end{equation}
where $\langle h \rangle=v=f\sqrt{\xi}\simeq246$ GeV and we work in the assumption of $v\ll f$.

\subsection{Composite Resonances}
\subsubsection{Fermion Resonances}\label{Fermion_Resonances}
In order to generate fermion Yukawa couplings and the effective potential of the composite Higgs and the composite DM, we adopt the partial compositeness scenario \cite{Kaplan:1991dc}. Additionally, when we formally embed the SM fermions in $\SO(6)$ representations, the embedding should preserve the $\mathbb{Z}_2$ symmetry stabilizing the DM. According to \cite{Frigerio:2012uc,Marzocca:2014msa}, we embed the left and right handed fermions in the fundamental representation of $\SO(6)$: 
\begin{align}
\xi^{u}_{L} =& \frac{1}{\sqrt{2}}{\left(\begin{array}{cccccc} b_L & -ib_L & t_L & it_L & 0 &0 \end{array}\right)}_{2/3}^T\,,\nonumber\\
\xi^{u}_{R}=&{\left(\begin{array}{cccccc} 0&0&0&0&0& t_R \end{array}\right)}_{2/3}^T\,,
\end{align}
where we focus on the top quark and the subscript is the $X$ charge assignment necessary to reproduce the top hypercharge. Other quarks and leptons can be embedded in a similar way, or could receive their mass from a different mechanism, as bilinear Yukawa-like interactions \cite{Ferretti:2014qta,Matsedonskyi:2014iha,Cacciapaglia:2015dsa}. Partial compositeness is introduced as
\begin{equation}
\mathcal{L} \simeq \epsilon \bar{\psi}_{SM} O_{\psi} +h.c\,.
\end{equation}
According to the CCWZ formalism, at low energy, $O_{\psi}$ can be represented as a function of $U(x)$ and $\Psi$, where $U$ is the NGB matrix and $\Psi$ is a collection of $\SO(5)$ fields. We focus for definiteness and for simplicity on cases of $\Psi$ resonances $S_i$ and $F_j$ transforming in the trivial and in the fundamental representation of $\SO(5)$. Details on the Lagrangian can be find in Appendix \ref{app: details}, where we also show how the effects of the heavy resonances can be encoded in form factors.

\subsubsection{Vector Resonances}
Vector resonances are generically expected as well as fermion resonances.
For simplicity we present one adjoint vector resonance $\rho_\mu$ and one fundamental vector resonance $a_\mu$, introduced following \cite{Contino:2011np}: again we refer to Appendix \ref{app: details} for detailed expressions.

\section{Dilaton extension of the composite Higgs model}\label{sec:dilaton extension}
As we previously stated the strong sector in isolation is a CFT enjoying a global $\SO(6)$ symmetry. In the vacuum both the conformal and the global symmetry are spontaneously broken. In this section we want to specify the general relations given in section \ref{sec: goldstone lagrangians} including the dilaton field. The dilaton dependence is introduced promoting $f$ to be a dynamical field $\chi=fe^{\sigma/f}$ and dressing composite fields with the appropriate powers of $\chi/f$. Notice that for simplicity we identify the scale associated to the dilaton $f_\sigma$ with $f$.
The Goldstone kinetic term becomes
\begin{align}
\mathcal{L}\supseteq&\frac{\chi^2}{4}\Tr[d_{\mu}d^{\mu}]\,.
\end{align}
In a similar manner the fermionic and vector Lagrangian 
are modified by the presence of the dilaton $\chi/f$.
At energies below the masses of the resonances the effective Lagrangian is
\begin{align}
\mathcal{L}_{eff}&\supseteq\Pi_{t_L}\bar{t}_L \slashed{p} t_L+\Pi_{t_{R}}\bar{t}_{R}\slashed{p}t_{R}-(\Pi_{t_{L}t_{R}}\bar{t}_{L}t_{R}+h.c)\nonumber\\
&+\frac{P^{\mu \nu}_{T}}{2}( \Pi_0 W_{\mu}^a W_{\nu}^a +\frac{\Pi_1 h^2}{4 f^2}(W^{1}_\mu W^{1}_\nu +W^{2}_\mu W^{2}_\nu))\nonumber\\
&+\frac{P^{\mu\nu}_{T}}{2}( \Pi_B B_\mu B_\nu +  \frac{\Pi_1h^2}{4 f^2 \cos^{2}\theta_w}Z_\mu Z_\nu )
\end{align}
where the form factors are modified by the presence of the dilaton.

The scalar potential $V(h,\eta,\chi)$ is obtained integrating out the SM top and vector bosons with a standard one loop computation. We briefly review the results of this computation in the following.

\subsection{The Scalar Potential}
The gauge contribution to the scalar effective potential $V_g(h,\eta , \chi)$ is
\begin{align}
V_{g}=&\frac{3}{2}\int\frac{d^{4} p_{E}}{(2 \pi)^4} (2 \log[\Pi_{WW}(-p_{E}^{2})]\\
&+\log[\Pi_{BB}(-p^{2}_{E})\Pi_{WW}(-p^{2}_{E})-\Pi_{W_{3}B}(-p^{2}_{E})])\nonumber\,.
\end{align}
Notice that as a result no potential is generated for $\eta$.
Fermion loops generate in principle all the possible terms containing Higgs and $\eta$ fields, but the case $N_F=N_S=1$ leads to the unsatisfactory prediction $\mu_{\eta}=\lambda_{\eta}=0$. Therefore we move to the next to minimal case, namely $N_F=1$, $N_S=2$.
The fermion contribution to the effective potential $V_f(h,\eta , \chi)$ is computed from
\begin{equation}
V_{f}=-2 N_{c}\int\frac{d^4 p_{E}}{(2\pi)^4}\log(p_{E}^2 \Pi_{t_{L}}\Pi_{t_{R}}+\Pi_{t_{L}t_{R}}^2)\,.
\end{equation}
We impose the generalized Weinberg sum rules \cite{Marzocca:2012zn}
and in order to get unsuppressed $\mu_{\eta}$ and $\lambda$, we assume $m_{2S}>> m_{F}>>m_{1S}\sim f$.

There is one subtlety: loops of top quarks, due to the large top Yukawa, induce a mixing between the Higgs and the dilaton field. Indeed the most general Lagrangian takes the form
\begin{equation}
\begin{split}
V_{f}(h,\eta,\chi)=& \frac{\chi^4}{f^4}\sum_{i+j <3}\kappa_{i,j} \frac{\chi^{2 \gamma(i+j)}}{f^{2\gamma(i+j)}} h^{2 i} \eta^{2 j}
\end{split}
\end{equation}
where $\gamma$ is the top anomalous dimension \cite{Chacko:2012sy}. Therefore
\begin{equation}
\frac{<\partial_{\chi}\partial_{h}V>}{<\partial_{h}^2 V>}\simeq \frac{\gamma v}{f}
\end{equation}
and we get that the mixing is proportional to the top anomalous dimension: since $\gamma\simeq0$ we safely neglect it. Similarly the Higgs radion mixing has been studied in a warped extra dimensional background and argued to be small for a pNGB Higgs \cite{Cox:2013rva}. We refer to Appendix \ref{app: dilaton} for a discussion on the dilaton potential. In the following we are going to treat the dilaton mass as a free parameter of the model, given its unpredictability in an effective description.

\subsection{Interactions with Massless Gauge Bosons}

The precise determination of interaction couplings between scalars such as dilaton, DM, and Higgs and gauge bosons is of primary importance in order to study LHC phenomenology and various aspects of DM detection. We therefore proceed in analyzing them.

First, we study the dilaton. It couples to gauge bosons via trace anomaly terms, which depend on the beta functions of the theory, and via triangle diagrams generated by loops of charged fields \cite{Kniehl:1995tn,Csaki:2007ns,Eshel:2011wz,Gillioz:2012se,Bellazzini:2012vz,Blum:2014jca}:
\begin{align}
\mathcal{L}\supseteq& \frac{\alpha_s}{8\pi}(b_{IR}^{3}-b_{UV}^{3}+\frac{1}{2}F_{1/2}(x_t))\frac{\sigma}{f}G_{\mu\nu}^{a}G^{a\mu\nu}\\
&+\frac{\alpha_{em}}{8\pi}(b_{IR}^{em}-b_{UV}^{em}+\frac{4}{3}F_{1/2}(x_t)-F_1 (x_W))\frac{\sigma}{f}F_{\mu\nu}F^{\mu\nu}\nonumber
\end{align}
where $x_i =4m^{2}_{i}/m^{2}_{\sigma}$. $F_{1/2}$ and $F_{1}$ are loop functions defined as
\begin{align}
F_{1/2}(x)&=2x(1+(1-x)f(x)),\\
F_1 (x)&=2+3x+3x(2-x)f(x),\nonumber\\
f(x)&=\begin{cases} \arcsin^{2}(1/\sqrt{x}) &\mbox{if } x \geq 1 \\ -\frac{1}{4}(\log(\frac{1+\sqrt{x-1}}{1-\sqrt{x-1}})-i \pi)^{2} &\mbox{if } x<1 \end{cases}\,.\nonumber
\end{align}
The loops of heavy top partners cancel with the IR beta function of the same in the limit of masses larger than $m_\sigma/2$, as we discuss in Appendix \ref{app: decoupling}. Therefore the top partners decouple and the only effects from the IR are from the light degrees of freedom.  

Among the light composite states we count the Higgs boson doublet, which enters the beta function coefficients with
\begin{equation}
b_{IR}^{2}=-\frac{1}{6}\,,\quad b_{IR}^{1}=-\frac{1}{6}\,.
\end{equation}
In case the right handed top is fully composite then
\begin{equation}
b_{IR}^{3}=-\frac{1}{3}\,,\quad b_{IR}^{1}=-\frac{8}{27}N_c\,,
\end{equation}
while it does not contribute to the composite beta functions if it is elementary.
As a result the IR beta function coefficients are 
\begin{equation}
b_{IR}^{3}\simeq 0\,,\quad b_{IR}^{em}=-\frac{1}{3}\,,
\end{equation}
or 
\begin{equation}
b_{IR}^{3}=-\frac{1}{3}\,,\quad b_{IR}^{em}=-\frac{11}{9}
\end{equation}
if also $t_R$ belongs to the composite fields.
The UV coefficients $b_{UV}^{3,em}$ are model dependent and we cannot specify them in our effective construction. Since they enter the couplings of the dilaton in the following discussion we will focus on simple benchmark values.

We now turn to Higgs couplings. According to \cite{Gillioz:2012se} the effect of composite fermion loops is expected to be negligible and the main contribution is given by top loops, closely resembling the SM result:

\begin{figure*}[t!]
	\includegraphics[width=\columnwidth]{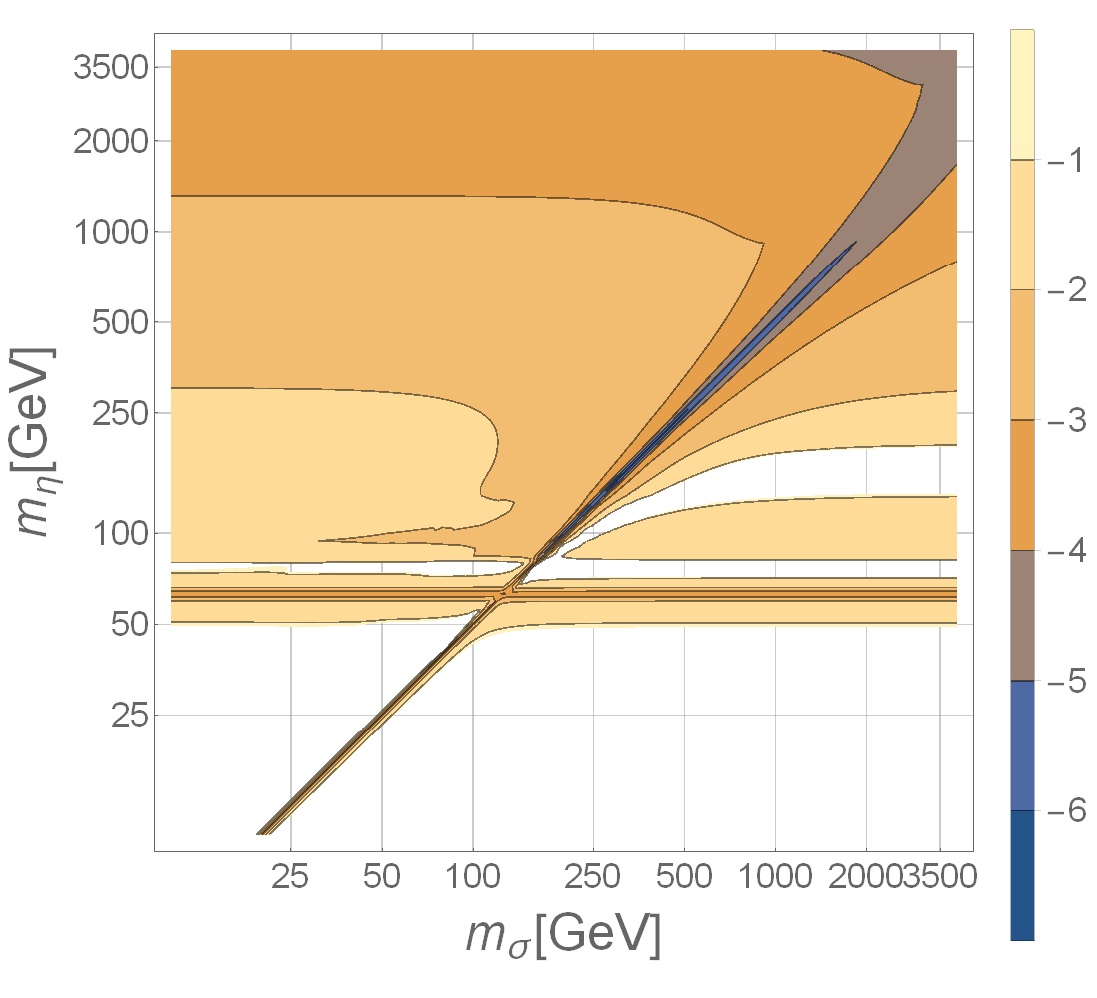}
	\includegraphics[width=\columnwidth]{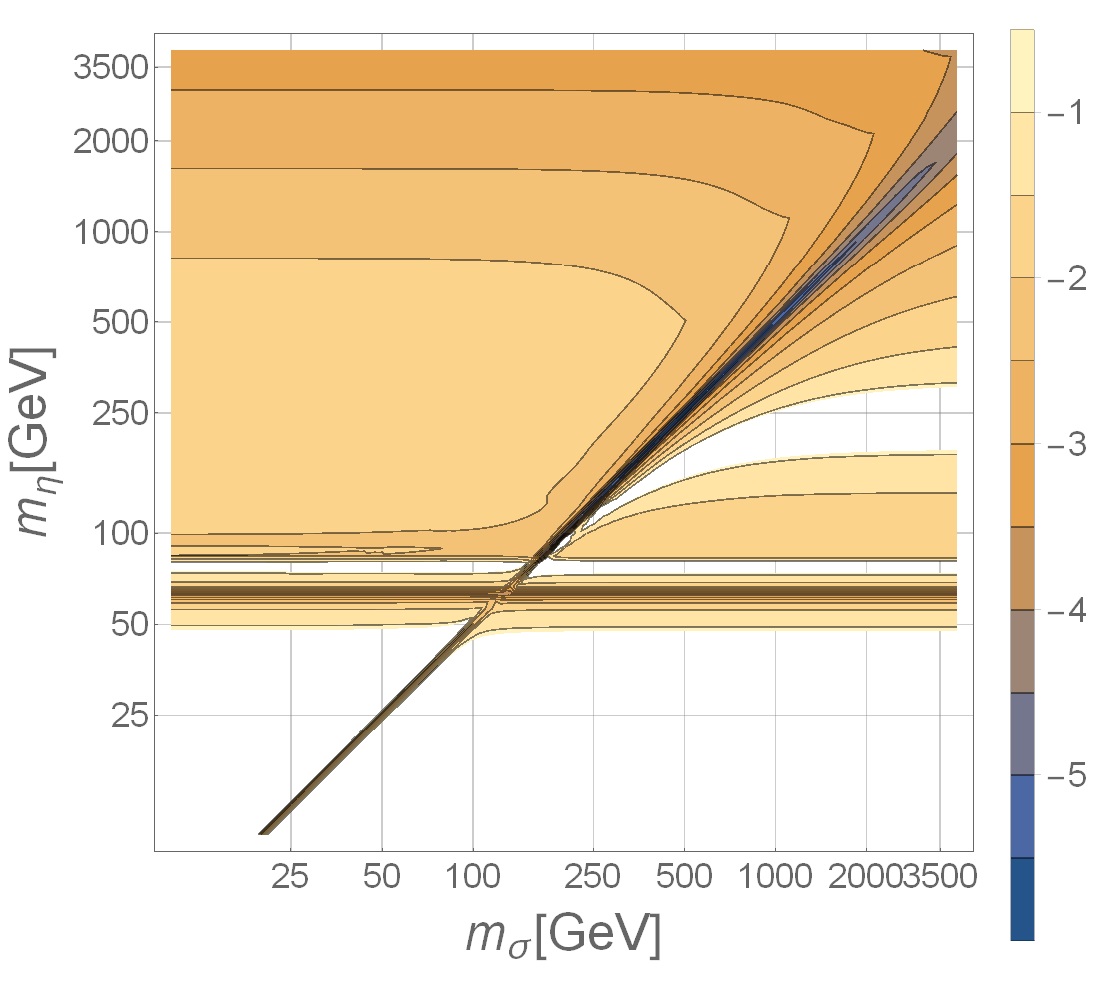}
	\caption{DM relic density  at $f=1000$ GeV (right) and $f=1500$ GeV (left). We contour $\log_{10}(\Omega h^2)<\log_{10}(0.12)$.}\label{Relic2_total}
\end{figure*}

\begin{align}
\mathcal{L}\supseteq &\frac{\alpha_{em}}{8\pi}(\frac{1-2\xi}{\sqrt{1-\xi}}4 q^{2}_{t}-\sqrt{1-\xi}F_1 (\frac{4m^{2}_{W}}{m^{2}_{h}}))\frac{h}{v}F_{\mu\nu}F^{\mu\nu}\nonumber\\
&+\frac{\alpha_s}{12\pi}\frac{1-2\xi}{\sqrt{1-\xi}}\frac{h}{v}G^{a}_{\mu\nu}G^{a\mu\nu}\,.
\end{align}

Similarly, since DM couples at tree level to SM fermions, we have DM to gauge bosons interactions at one loop. Given the coupling of $\eta$ to fermions \footnote{The couplings of $\eta$ to SM fermions other than top depend on the formal embedding of the SM quarks into $\SO(6)$ representations. We fix for convenience the same couplings for all the quarks.}
\begin{equation}
\mathcal{L}\supseteq \frac{\xi}{2(1-\xi)}m_{\psi}\bar{\psi}\psi \frac{\eta^2}{v^{2}}
\end{equation}
we easily read the couplings to gauge bosons
\begin{align}
\mathcal{L}\supseteq&-\frac{\alpha_s}{32\pi}F_{1/2}(\frac{4m^{2}_{t}}{m^{2}_{\eta}})\frac{\xi}{1-\xi}\frac{\eta^{2}}{v^2}G^{a}_{\mu\nu}G^{a\mu\nu}\nonumber\\
&-\frac{3\alpha_{em}}{16\pi}\frac{\xi}{1-\xi}\frac{\eta^2}{v^2}q^{2}_{t}F_{1/2}(\frac{4m^{2}_{t}}{m^{2}_{\eta}})F_{\mu\nu}F^{\mu\nu}\,.
\end{align}
We neglect possible couplings of $\eta$ to pair of gauge bosons arising from the Wess-Zumino-Witten term, they could be computed in principle given the details of the fundamental underlying theory, as done in \cite{Arbey:2015exa}.

\subsection{Effective Lagrangian}
An effective Lagrangian for the SM fields, the DM candidate $\eta$ and the dilaton $\sigma$ is obtained, expanding the scalars around their VEV
\begin{align}
h&= v+\sqrt{1-\xi }h_{phys}\,,\,\,
\eta=\eta_{phys}\,,\,\,
\sigma=\sigma_{phys}\,.
\end{align}
The resulting Lagrangian, the starting point of our phenomenological analysis, has the following form:
\begin{align}\label{eq:pheno Lag}
\mathcal{L}\supseteq& +\frac{1}{2}(\partial_\mu h)^2 (1+a_{hh} \frac{h}{v}+b_{hh}\frac{h^2}{v^2}+b_{h\eta}\frac{\eta^2}{v^2}) e^{2\sigma/f} \nonumber\\
&+ \frac{1}{2}(\partial_\mu \eta)^2 (1+b_{\sigma \eta} \frac{\eta^2}{v})e^{2\sigma/f}\nonumber\\
& +(\partial_\mu h \partial^\mu \eta)(c_{\eta}\frac{\eta}{v}+d_{\eta h}\frac{\eta h}{v^2})e^{2\sigma/f}\nonumber\\
&+\sum_{V}\frac{m^{2}_{V}}{2}V^{\mu}V_{\mu}(1+\sqrt{1-\xi}\frac{h}{v})^2 e^{2\sigma/f}\nonumber\\
&-\sum_{i}m_{\psi}\bar{\psi_i }\psi (1+a_{\psi h}\frac{h}{v}+b_{\psi h} \frac{h^2}{v^2})e^{\sigma/f}\nonumber\\
&+ \frac{\alpha_s}{8\pi}(b_{IR}^{3}-b_{UV}^{3}+\frac{1}{2}F_{1/2}(x_t))\frac{\sigma}{f}G_{\mu\nu}^{a}G^{a\mu\nu}\nonumber\\
&+\frac{\alpha_{em}}{8\pi}(b_{IR}^{em}-b_{UV}^{em}+\frac{4}{3}F_{1/2}(x_t)-F_1 (x_W))\frac{\sigma}{f}F_{\mu\nu}F^{\mu\nu}\nonumber\\
&+\frac{\alpha_{em}}{8\pi}(\frac{1-2\xi}{\sqrt{1-\xi}}4 q^{2}_{t}-\sqrt{1-\xi}F_1 (\frac{4m^{2}_{W}}{m^{2}_{h}}))\frac{h}{v}F_{\mu\nu}F^{\mu\nu}\nonumber\\
&+\frac{\alpha_s}{12\pi}\frac{1-2\xi}{\sqrt{1-\xi}}\frac{h}{v}G^{a}_{\mu\nu}G^{a\mu\nu}\nonumber\\
&-\frac{\alpha_s}{32\pi}F_{1/2}(\frac{4m^{2}_{t}}{m^{2}_{\eta}})\frac{\xi}{1-\xi}\frac{\eta^{2}}{v^2}G^{a}_{\mu\nu}G^{a\mu\nu}\nonumber\\
&-\frac{3\alpha_{em}}{16\pi}\frac{\xi}{1-\xi}\frac{\eta^2}{v^2}q^{2}_{t}F_{1/2}(\frac{4m^{2}_{t}}{m^{2}_{\eta}})F_{\mu\nu}F^{\mu\nu}\nonumber\\
&-V_{eff}(h,\eta,\chi)\,.
\end{align}
with $V_{eff}(h,\eta,\chi)$ given in \refeq{eq:Veff dilaton}.
Concerning the dilaton mass we are mostly interested in $m_\sigma > 0.1f$ \cite{Bellazzini:2012vz,Bellazzini:2013fga,Coradeschi:2013gda}, because a too light dilaton requires too much fine tuning, and $m_{\sigma} \leq 4\pi f$ because of NDA \cite{Georgi:1992dw}.

\section{Relic Abundance}\label{sec: relic abundance}
\subsection{Introduction to WIMPs}
WIMP is one of the most compelling paradigm for DM. In case of scalar DM fundamental and composite singlet scalar WIMPs have been extensively studied, see e.g. \cite{Guo:2010hq,Frigerio:2012uc,Cline:2013gha,Marzocca:2014msa}.

In order to implement the WIMP scenario, we need to assume that the DM candidate is in thermal equilibrium since the very early universe. In case of composite DM there exists an energy threshold above which DM particles are resolved in their constituents. Since we have $f\gg v$ we can safely assume thermal equilibrium; moreover heavy degrees of freedom of the strong theory are irrelevant being, indeed, heavy. As a result we can use the standard picture of WIMPs \cite{Lee:1977ua}.

We recall that the measured DM relic density is $\Omega h^2 =0.1199\pm0.002$ \cite{Ade:2013zuv}. The current relic density is predicted using the Weinberg-Lee equation \cite{Lee:1977ua}
\begin{equation}
\frac{dn}{dt}+3Hn=<\sigma v> (n^{2}_{eq}-n^{2})
\end{equation}
where $\langle\sigma v\rangle$ is the thermal average of cross sections times relative speed, and $H$ is the Hubble constant. Expanding $\sigma v$ for small velocities as $\sigma v= a+ bv^2$ we get $<\sigma v> =a+ 6b/x$, where $x=m/T$. We use this expansion because $s$-wave processes are dominant in our model. By solving the above equation, we get the freeze out temperature
\begin{equation}
x_{F}=\ln\left(\frac{5}{4}\sqrt{\frac{45}{8}}\frac{g}{2\pi^3}\frac{M_{pl}m_{\eta}(a+6b/x_{F})}{\sqrt{g_* x_F}}\right)\,,
\end{equation}
where $g$ is the number of degrees of freedom of the DM and $g_*$ is the effective relativistic degrees of freedom in thermal equilibrium. 

As a result, the DM relic abundance is given by
\begin{equation}
\Omega h^2 \simeq \frac{1.07 \times 10^9}{\mbox{GeV } M_{pl}\sqrt{g_*}}\frac{x_F}{a+3(b-a/4)/x_F}\,.
\end{equation}

\subsection{Annihilation Cross Sections}
\begin{figure}[t!]
\centering
\includegraphics[width=\columnwidth]{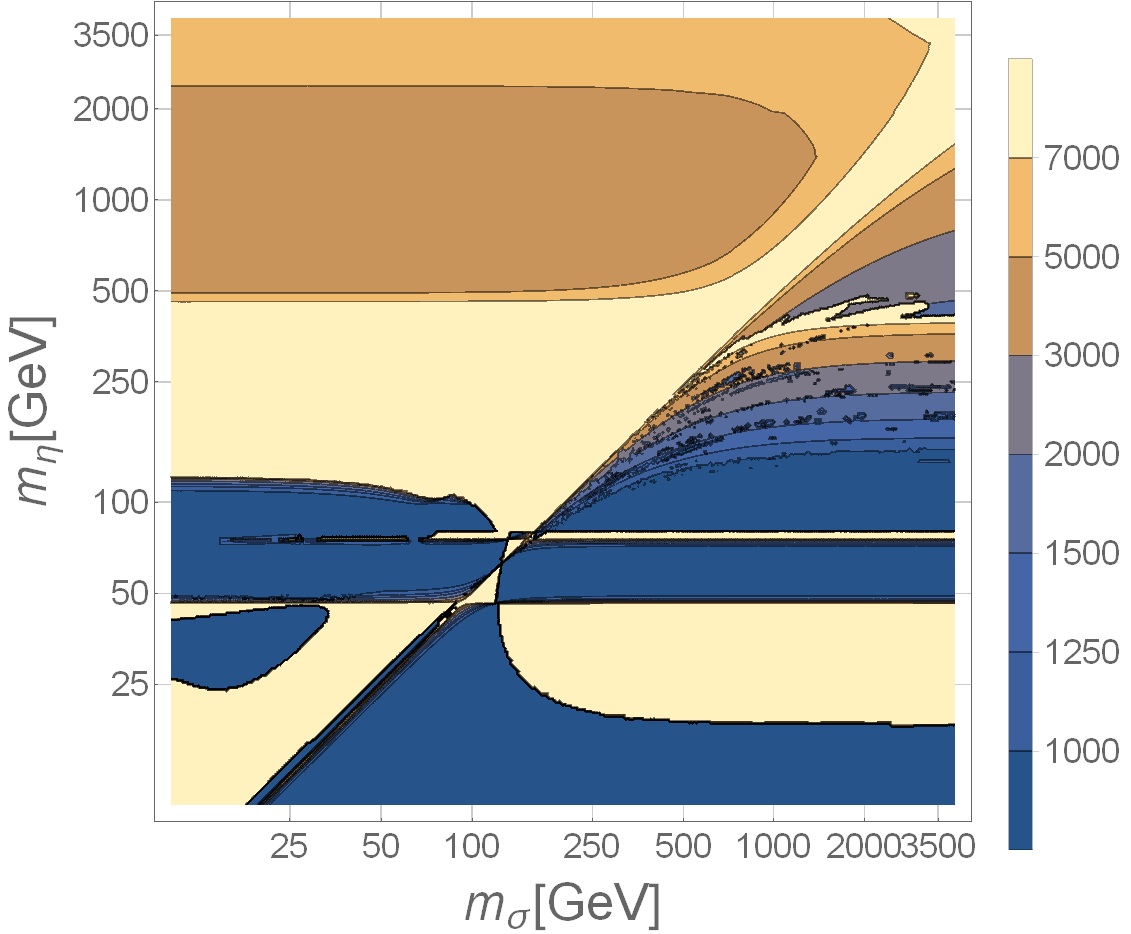}
\caption{Contour of values of $f$ in GeV necessary to reproduce the observed relic density.}
\label{f_relic}
\end{figure}
In our model the DM candidate is the fifth pseudo Goldstone boson of the coset $\SO(6)/\SO(5)$, $\eta$. Its effective potential is determined by the underlying theory and can be reliably computed using an effective IR Lagrangian, as we outlined before: the form of this Lagrangian depends on the details of the theory, as the number of top partners $N_F$ and $N_S$. If $N_F=N_S=1$ the mass is fixed to be $m_\eta\simeq m_h/2$ and the predicted relic density is too small to be a viable option. Therefore we focus on the next to minimal case $N_F=1$, $N_S=2$, where the $\eta$ mass varies as a free parameters over an interval. We fix the portal coupling $\lambda_{h\eta}\simeq0.13$, following \cite{Marzocca:2014msa}. 

We computed the annihilation channels including $\eta \eta\rightarrow WW, ZZ, hh, h\sigma, \sigma\sigma, AA, GG$, and $\bar{\psi}\psi,$ where $\psi$ runs over the SM fermions. Note that the above processes are dominated by $s$-wave exchange since $p$ and higher order terms are suppressed by $v^2$. Full expressions are reported in Appendix \ref{appendix:annihilation}. We present here asymptotic forms valid in certain limits. We focus on $m_{\sigma}, m_\eta \gg m_Z$: as a result $\eta\eta\rightarrow VV$ dominates the annihilation cross section.

First we take $m_\eta\gg m_\sigma$. If this is the case we obtain
 \begin{equation}\label{eq: param cross section}
 \langle\sigma v_{AA}\rangle\simeq\frac{m_{\eta}^{2}}{c_{AA} \pi f^4}\,,
 \end{equation}
where $c_{ZZ}=16$, $c_{WW}=8$, $c_{\sigma\sigma}=4$ and $c_{hh}=16$. The $\eta\eta\rightarrow h\sigma$ process is controlled by $\eta\partial_\mu h\partial^\mu \eta$ and suppressed by $\xi^{3}/(1-\xi)$.

\begin{figure*}[!t]
	\includegraphics[width=\columnwidth]{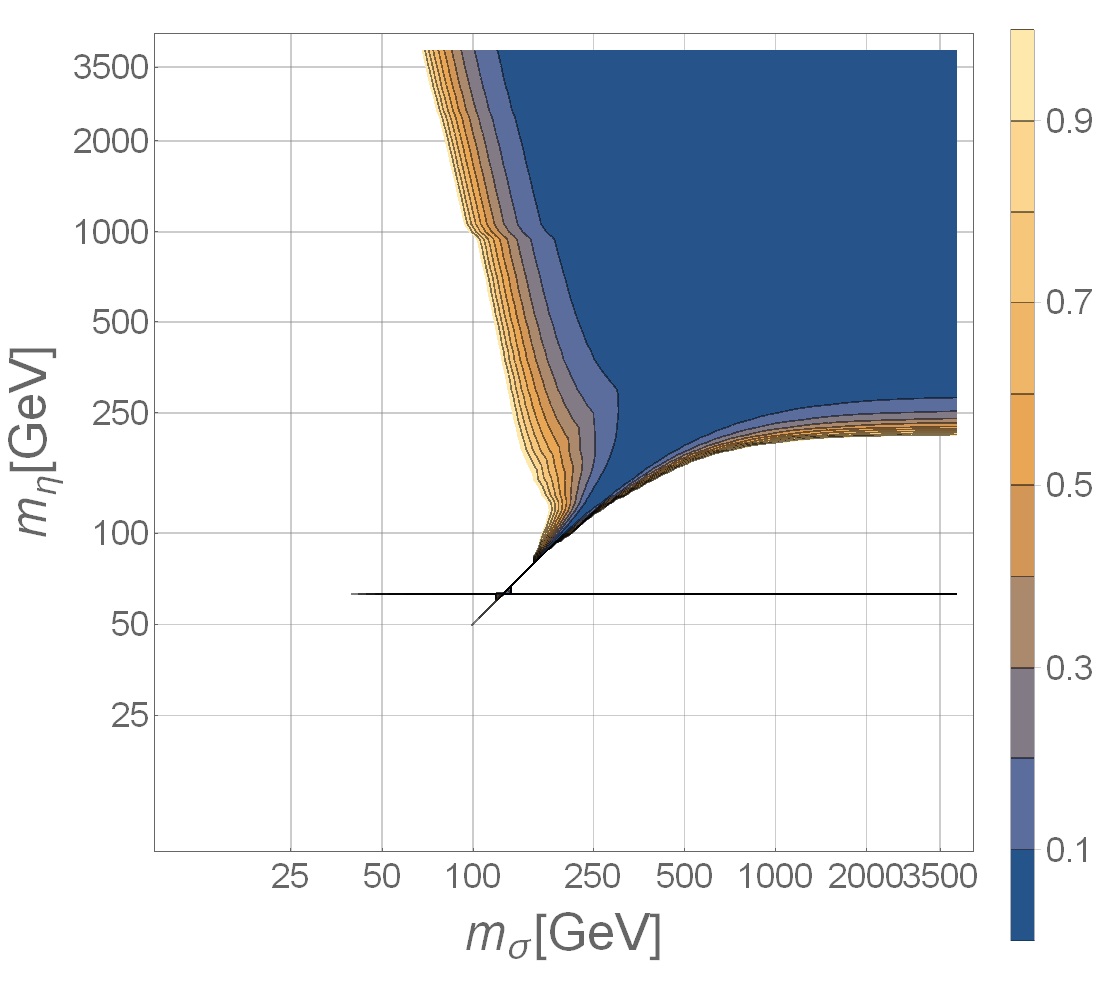}
	\includegraphics[width=\columnwidth]{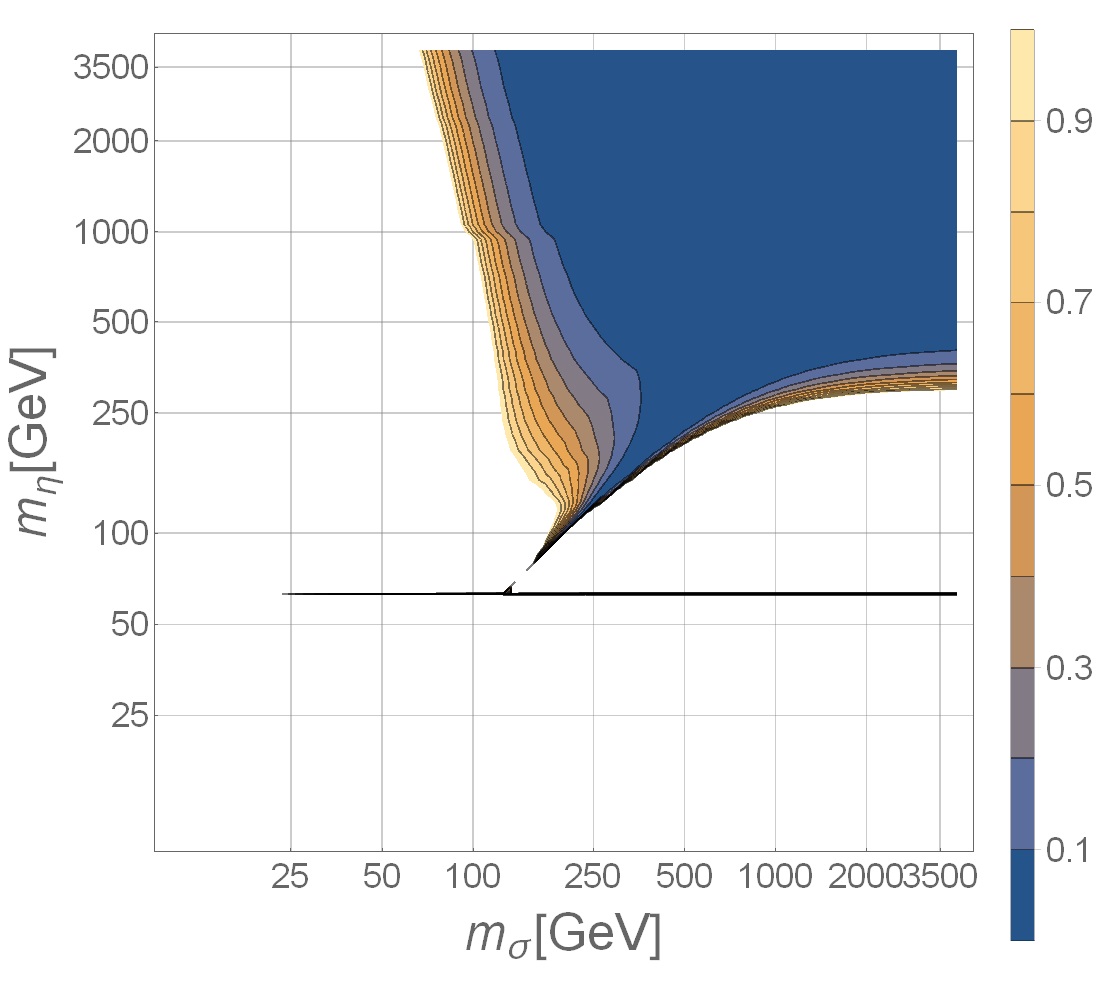}
	\caption{Allowed region at $90\%$ confidence level with $f=1000$ GeV (left) and $f=1500$ GeV (right) from direct searches. We contour the ratio of nucleon-DM cross section over LUX cross section bound.}\label{LUX_total}
\end{figure*}

The total thermally averaged cross section is then
\begin{equation}
\langle\sigma v \rangle\simeq \frac{m^{2}_{\eta}}{2 \pi f^4} \simeq 3\times10^{-26}{\left(\frac{8.5 \text{ TeV}\,m_\eta}{f^2}\right)}^{2}\text{cm}^3 /\text{s}\,.
\end{equation}
Note that $\langle \sigma v\rangle$ should be equal to or larger than $3\times 10^{-26} \text{cm}^3/\text{s}$  in order to reproduce a relic density equal to or smaller than the observed one.
 
In the massive dilaton limit, $m_{\sigma}\gg m_{\eta}$, the dilaton exchanging processes are suppressed by $m^{2}_{\eta}/m^{2}_{\sigma}$ and Higgs exchanging processes have a similar asymptotic form as before.
Consequently we get larger annihilation cross section, parametrized as in \refeq{eq: param cross section}
where now $c_{ZZ}=4$, $c_{WW}=2$, and $c_{hh}=4$. As a result, the total thermally averaged cross section is
\begin{equation}
\langle \sigma v\rangle \simeq \frac{m^{2}_{\eta}}{\pi f^4}\simeq 3\times 10^{-26}{\left(\frac{12 \text{ TeV}\,m_\eta}{f^2}\right)}^{2} \text{cm}^{3}/\text{s}\,.
\end{equation}
In Fig.~\ref{Relic2_total} we present the predicted relic density of DM particles in the $m_\sigma-m_\eta$ plane.
We clearly distinguish a depletion of $\Omega h^2$ in correspondence of the points with $m_\eta=m_h/2\simeq63$ GeV and $m_\eta=m_\sigma/2$. If Fig.~\ref{f_relic} we present the value of the scale $f$ which is necessary to reproduce the observed relic density, in the same plane.

\section{Direct Detection}\label{sec: direct detection}
Null results from direct detection experiments, as LUX \cite{Akerib:2013tjd,Akerib:2015rjg}, put limits on the nucleon-DM scattering cross section. The interactions in \refeq{eq:pheno Lag} relevant in this regard are the vertices between the scalars $h$, $\eta$ and $\sigma$ with the fermion bilinears $\bar\psi\psi$ and the field strength operator $G_{\mu\nu}G^{\mu\nu}$ of colored interactions.
From those we derive an effective theory for nucleons
\begin{align}
\mathcal{L}\supseteq& - \sum_{i=n,p}\bar{\psi}_i \psi_i (y_{s,i} \sigma + y_{h,i} h +y_{\eta,i} \eta^2 )\,,
\end{align}
where
\begin{align}
y_{\sigma,i} =&\sum_\psi \frac{1}{f} \langle i| m_\psi \bar{\psi}\psi| i\rangle-\frac{C_s}{8\pi f}\langle i|\alpha_s G^{a}_{\mu\nu}G^{a\mu\nu}|i\rangle\,,\nonumber\\
y_{h,i}=&\frac{1}{v}\frac{1-2\xi}{\sqrt{1-\xi}} (\sum_\psi\langle i|m_\psi \bar{\psi}\psi |i\rangle -\frac{1}{12\pi}\langle i|\alpha_s G^{a}_{\mu\nu}G^{a\mu\nu}|i\rangle)\,,\nonumber\\
y_{\eta,i}=&-\frac{1}{2v^2}\frac{\xi}{1-\xi}\sum_\psi\langle i|m_\psi \bar{\psi}\psi|i\rangle\\
&+\frac{1}{32 \pi^{2}v^2}\frac{\xi}{1-\xi}F_{1/2}(\frac{4m^{2}_{t}}{m^{2}_{\eta}})\langle i|\alpha_s G^{a}_{\mu\nu}G^{a\mu\nu}|i\rangle\,,\nonumber
\end{align}
where $i$ stands for neutron and proton and $\psi$ stands for SM quarks \footnote{The couplings of $\eta$ and $h$ to SM fermions other than top depend on the formal embedding of the SM quarks into $\SO(6)$ representations. We fix for convenience the same couplings for all the quarks. Small changes should not make a difference our analysis.}. Integrating out the dilaton and the Higgs we obtain
\begin{equation}
\begin{split}
\mathcal{L}_{eff}\supseteq - a_{n} \bar{n} n \eta^2- a_{p} \bar{p}p \eta^2
\end{split}
\end{equation}
where
\begin{equation}
a_{i}\simeq y_{\eta,i}-\frac{2 m^{2}_{\eta} y_{\sigma,i}}{f m^{2}_{\sigma}}-\frac{\lambda_{h\eta} v \sqrt{1-\xi} y_{h,i}}{2 m^{2}_{h}}\,.
\end{equation}
For the matrix elements, we take the values for $u$ and $d$ quarks from \cite{Bishara:2015cha}, and for $s$, $c$, $b$, and $t$ quarks from \cite{Junnarkar:2013ac}:
\begin{align}
&f^{i}_{\psi}=\langle i|\bar{\psi}\psi|i\rangle \frac{m_{\psi}}{m_{i}}\nonumber
\\ &f^{n}_{u}\simeq 0.016\,, f^{p}_{u}\simeq 0.018\,,
 f^{n}_{d}\simeq 0.038\, , f^{p}_{d}\simeq 0.034\,,\nonumber
\\ &f^{n}_{s}\simeq f^{p}_{s}\simeq 0.043\,,
 f_{c} \simeq 0.0814\,,
 f_{b} \simeq 0.0785\,,
f_{t} \simeq  0.0820\,,\nonumber
\\ &\alpha_s \langle  n|G^{a}_{\mu\nu}G^{a\mu\nu}|n\rangle \simeq -2.4\text{ GeV}\,.
\end{align}
We then derive the nucleon-DM cross section
\begin{equation}
\sigma_{\eta ,i}\simeq \frac{ a_{i}^2 m^{2}_{i}}{\pi m^{2}_{\eta}}\,.
\end{equation}
By comparing with the LUX data we get the allowed parameter region, shown in Fig.~\ref{LUX_total}. For the points for which the model predicts a relic density lower than the observed one we rescale the bound.

\section{Indirect Detection}\label{sec: indirect detection}
\subsection{Sommerfeld Enhancement}
\begin{figure*}[t]
	\includegraphics[width=\columnwidth]{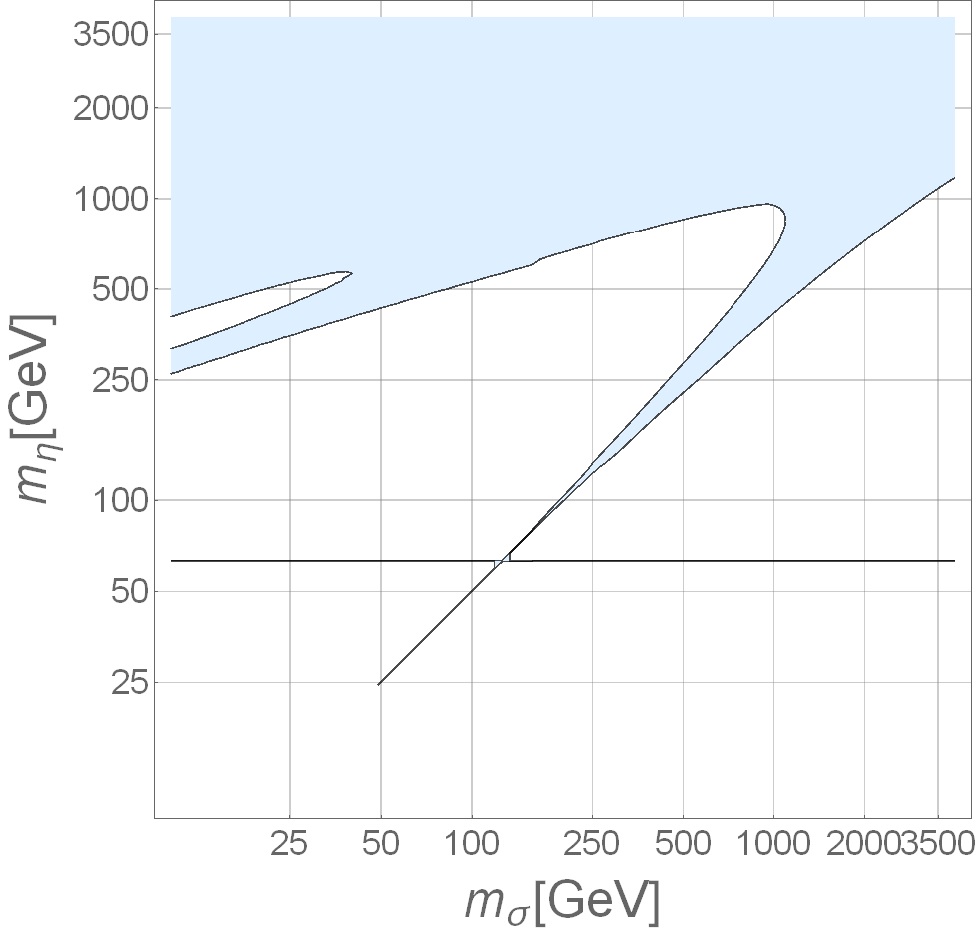}
	\includegraphics[width=\columnwidth]{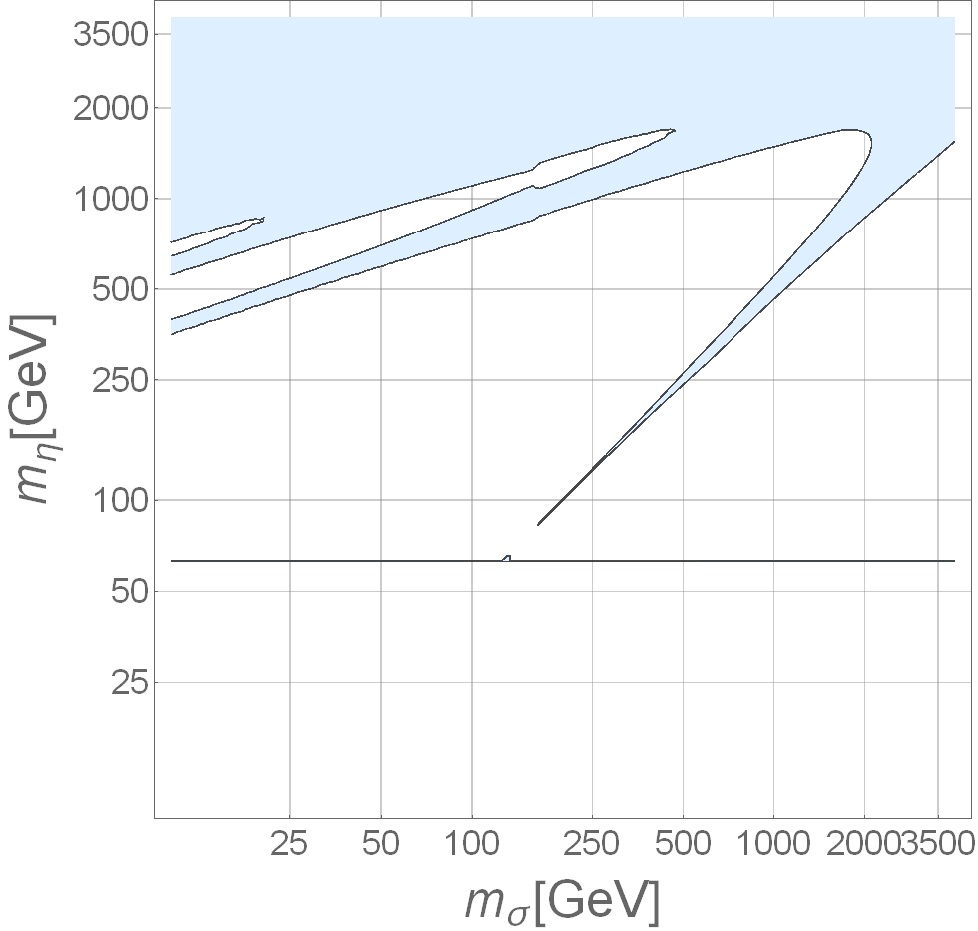}
	\caption{Excluded parameter region with $f=1000$ GeV (left) and $f=1500$ GeV (right), using the informations on antiproton fluxes from the Galactic gas.}\label{Ex_total}
\end{figure*}
To correctly evaluate the signals searched by indirect detection experiments we take into account Sommerfeld enhancement, following \cite{Iengo:2009ni,Cassel:2009wt}. To this end we need the three fields interaction vertices of DM $\eta$ with dilaton and Higgs, which are respectively of the form
\begin{equation}
\frac{1}{2}(\partial \eta)^{2} 2 \frac{\sigma}{f}- \frac{1}{2}m^{2}_{\eta} \eta^{2} 4\frac{\sigma}{f}
\end{equation}
and
\begin{equation}
\frac{\xi}{1-\xi}\partial_\mu \eta \partial^\mu h \frac{\eta}{v}-v\sqrt{1-\xi}\frac{\lambda_{h\eta}}{2} \eta^{2}h
\end{equation}
where $\lambda_{h \eta}=0.013$. These lead to interaction potential, in momentum space, of the form
\begin{equation}
\begin{split}
V(p-q)=&\frac{1}{4m^{2}_{\eta}}(\frac{2}{f})^{2}\Pi_i \frac{(p_{i \mu}(p_0 +(-1)^{i+1}q )^{\mu}+2m^{2}_{\eta})}{(p-q)^{2}-m^{2}_{\sigma}-\Pi_{\sigma}(4m^{2}_{\eta}))}\\
\end{split}
\end{equation}
and
\begin{equation}
V(p-q)=\frac{1}{4m^{2}_{\eta}}\frac{\Pi_{i=1}^{2}(\frac{(-1)^{i}\xi}{v(1-\xi)}p_i (p-q) -v\sqrt{1-\xi}\frac{\lambda_{h\eta}}{2})}{(p-q)^{2}-m^{2}_{h}}
\end{equation}
for dilaton and Higgs respectively, where $p_1$ and $p_2$ are the momenta of the incoming particles and $p=(p_1-p_2)/2$. In the non-relativistic limit, in the instant interaction limit and in the CM frame the above expressions reduce to
\begin{equation}
V(p-q)=-\frac{1}{4m^{2}_{\eta}}(\frac{2}{f})^{2} \frac{m^{4}_{\eta}}{(\vec{p}-\vec{q})^{2}+m^{2}_{\sigma}}
\end{equation} 
and 
\begin{equation}
V(p-q)=-\frac{1}{4m^{2}_{\eta}}\frac{v^{2}(1-\xi) \lambda_{h\eta}^{2}}{4(\vec{p}-\vec{q})^{2}+4m^{2}_{h}}\,.
\end{equation}
As a result, the following Yukawa potential arises
\begin{equation}
V(r)=-\frac{\alpha_\sigma}{r}e^{-m_{\sigma} r}-\frac{\alpha_h}{r}e^{-m_{h}r}
\end{equation}
where $\alpha_\sigma = \frac{9m^{2}_{\eta}}{4\pi f^{2}}$ and $\alpha_h=\frac{ (1-\xi)v^{2}\lambda_{h\eta}^{2}}{16\pi m^{2}_{\eta}}$. Notice that $\alpha_{\sigma}\gg\alpha_{h}$, and DM is in non relativistic regime, thus Sommerfeld enhancement is dilaton dominated. According to \cite{Cassel:2009wt,Feng:2010zp,Slatyer:2009vg}, an analytic approximate formula for dilaton mediated Sommerfeld enhancement is
\begin{equation}\label{eq:Sommerfeld}
S=\frac{\pi}{\epsilon_v}\frac{\sinh(\frac{12 \epsilon_v }{\pi \epsilon_{\sigma}})}{\cosh(\frac{12 \epsilon_v }{\pi \epsilon_{\sigma}})-\cos(2\pi \sqrt{\frac{6}{\pi^{2} \epsilon_{\sigma}}-(\frac{6 \epsilon_{v}}{\pi^{2} \epsilon_{\sigma}})^{2}})}
\end{equation}
where $\epsilon_v =v/\alpha_\sigma$ and $\epsilon_\sigma =m_{\sigma}/(\alpha_{\sigma} m_{\eta})$. 

\subsection{Antiproton Flux}
\begin{figure}[t]
\centering
	\includegraphics[width=\columnwidth]{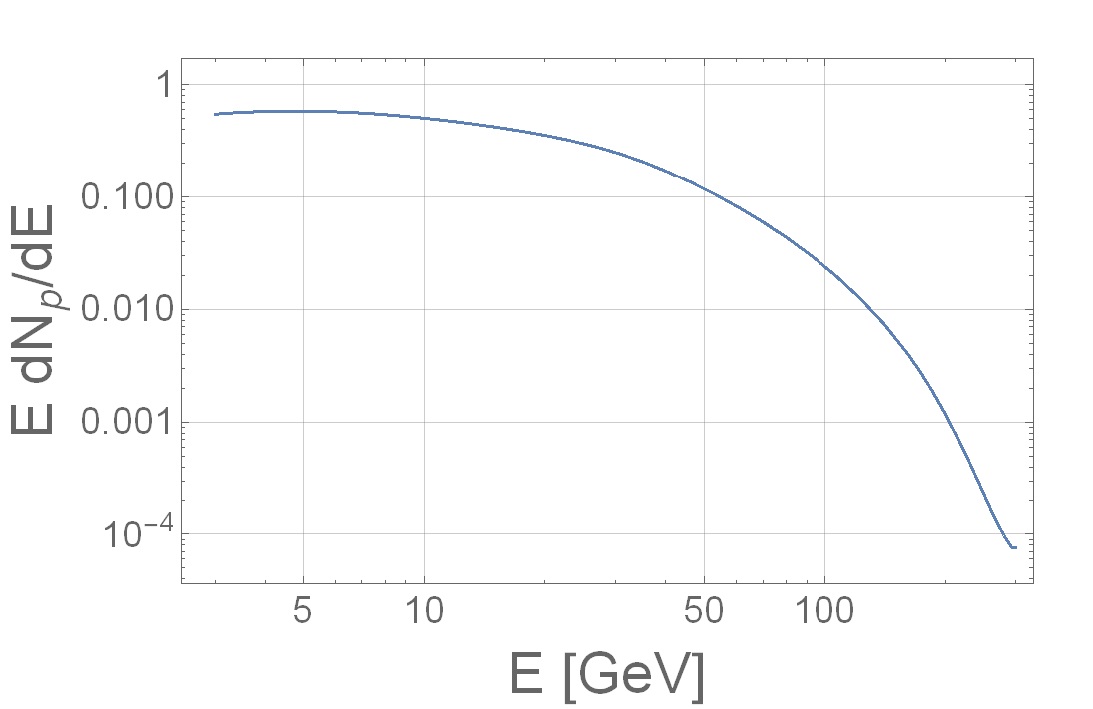}\caption{Differential antiproton spectrum per DM annihilation, computed for $m_{\eta}=300$ GeV, $m_{\sigma}=1000$ GeV and $f=1500$ GeV.}\label{Antiproton}
\end{figure}
\begin{figure*}[t!]
\centering
\includegraphics[width=\columnwidth]{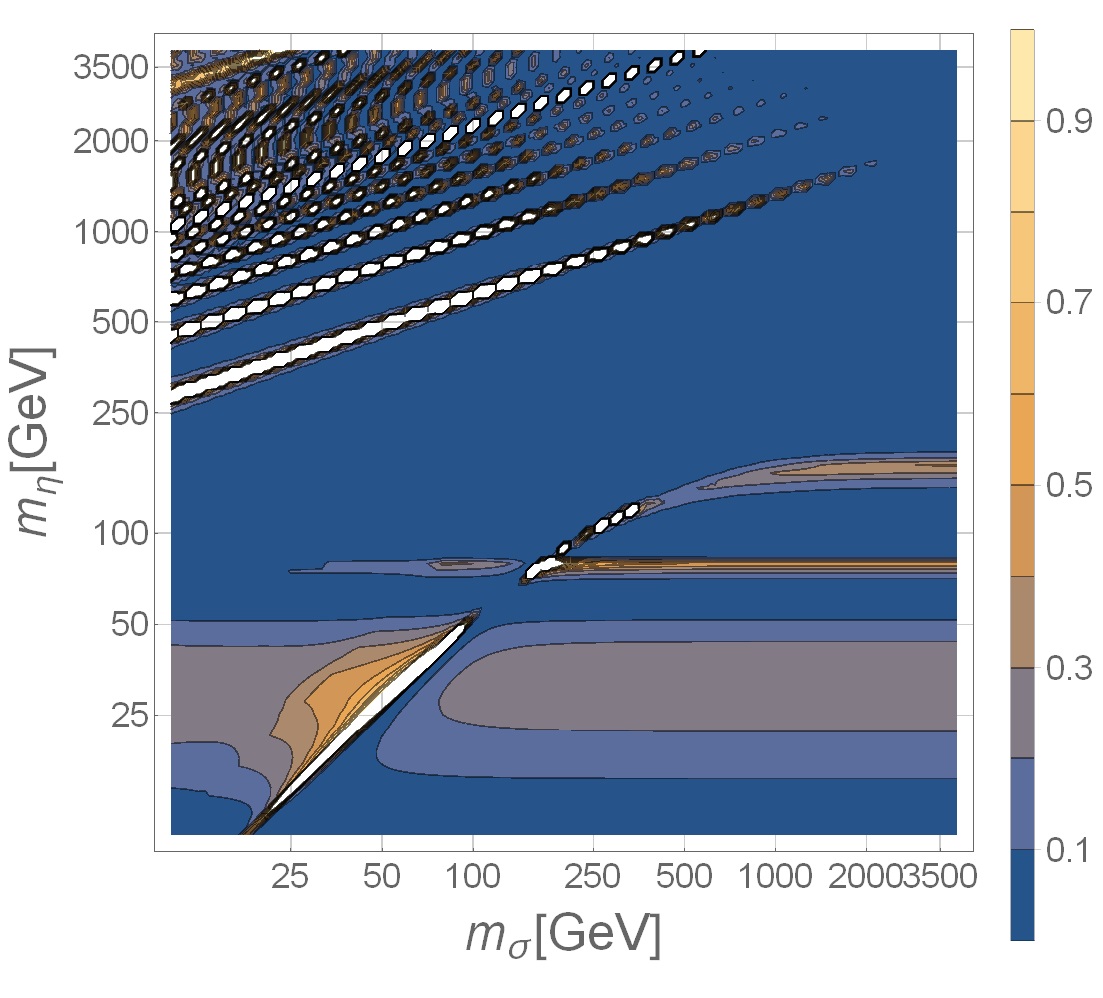}
\includegraphics[width=\columnwidth]{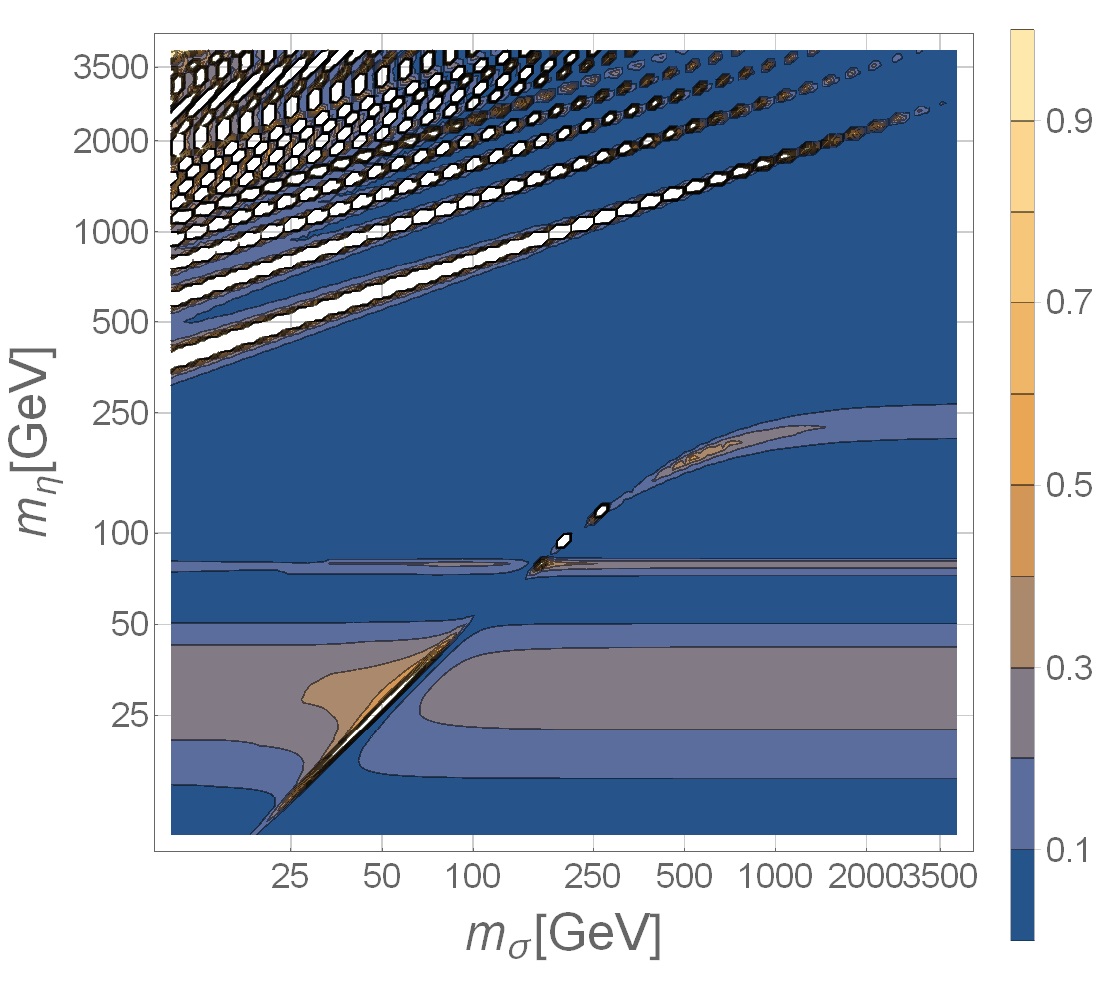}
\caption{$\bar{p}/p$ flux model prediction over the AMS results, computed at $f=1000$ GeV (left) and $f=1500$ GeV (right).}\label{pbarp_tot}
\end{figure*}
\begin{figure}[t!]
\centering
\includegraphics[width=\columnwidth]{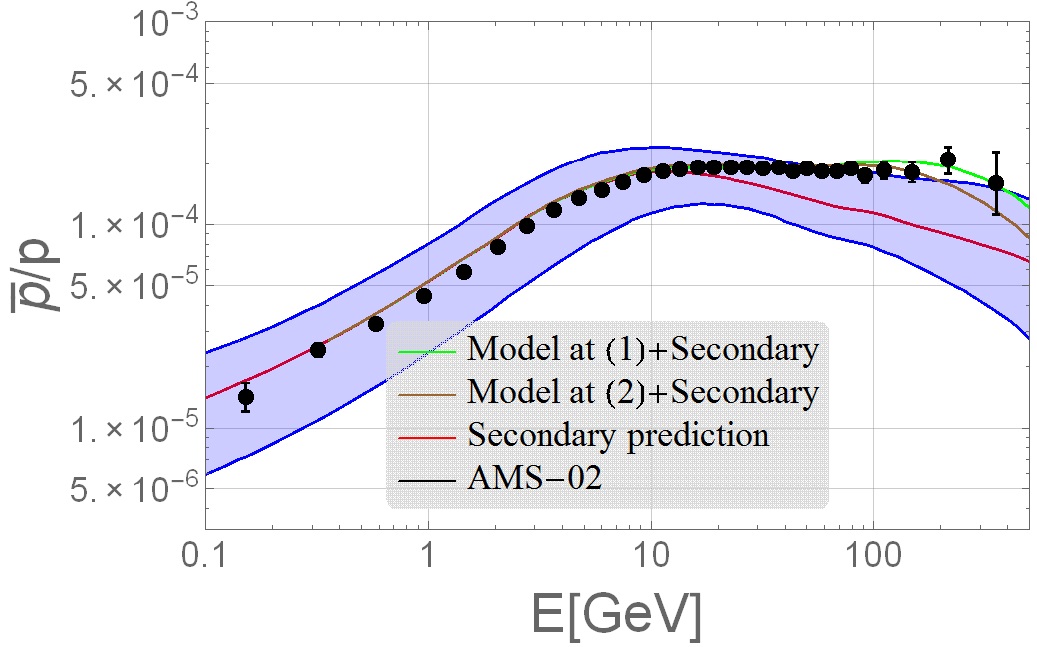}\caption{Antiproton to proton flux ratio. The data points are AMS-02 data, and the red line is the secondary prediction.
}\label{pbarp}
\end{figure}
DM annihilation can produce antiprotons in various ways and we take into account the AMS-02 \cite{Aguilar:2015ooa,Giesen:2015ufa} measure to constraint the parameter space of the model, demanding that the predicted antiproton flux does not exceed the observed one.
Following \cite{Blum:2014jca,Agashe:2009ja}, we derive a bound on the antiproton flux produced by DM annihilation by imposing that the amount of antiprotons produced by the DM annihilation in the Galactic disk is smaller than the antiproton flux due to primary cosmic rays colliding with interstellar medium in the disc  \cite{Katz:2009yd}.

We followed \cite{Ciafaloni:2010ti,Cirelli:2010xx} to compute antiproton spectrum, and \cite{Elor:2015tva} to evaluate cascade annihilation processes initiated by $\eta\eta\rightarrow\sigma\sigma\,,hh$, including the Sommerfeld enhancement \refeq{eq:Sommerfeld}.

The injection rate density of antiprotons produced by DM annihilation is
\begin{align}
Q_{\bar{p}}(E)=&\frac{1}{2}n^{2}_{\eta}\langle\sigma v\rangle \frac{dN_{\bar{p}}}{dE}
\\
\simeq& 5\times10^{-36} \text{cm}^{-3}\text{s}^{-1}\text{GeV}^{-1}\left(\frac{\rho_{\eta}}{0.4 \text{ GeV}\text{cm}^{-3}}\right)^{2}\nonumber\\
&\left(\frac{\langle \sigma v\rangle}{3\times 10^{-26}\text{ cm}^{3}\text{s}^{-1}}\right)\left(\frac{m_{\eta}}{1 \text{ TeV}}\right)^{-3}\left(m_{\eta}\frac{dN_{\bar{p}}}{dE}\right)\nonumber
\end{align}
where $\rho_{\eta}=m_{\eta}n_{\eta}$ and $dN_{\bar{p}}/dE$ is the differential antiproton spectrum per annihilation event. According to \cite{Elor:2015tva} dilaton and higgs contributions to antiproton flux is given by
\begin{equation}
\frac{dN_{\bar{p}}}{dx}=2\int^{t_{max}}_{t_{min}} \frac{dx_0}{x_0 \beta_{\sigma}}\frac{dN_{\bar{p},S}}{dx_0}
\end{equation}
where $S= h, \sigma$, $\beta_{\sigma}=\sqrt{1-\gamma^{-2}_{\sigma}}$, $x=E/m_{\eta}$, $t_{min}=2 x \gamma^{2}_{\sigma}(1- \beta_{\sigma})$, $t_{max}= \min[1,2 x \gamma^{2}_{\sigma}(1+\beta_{\sigma})]$ and $\gamma_{\sigma}=m_{\eta}/m_{\sigma}$. By including cascade effects, we obtain the full differential antiproton spectrum, following \cite{Cirelli:2010xx}. Fig.~\ref{Antiproton} shows a typical spectrum at $f=1500$ GeV and for $m_{\eta}=300$ GeV and $m_{\sigma}=1000$ GeV.

In order to impose our condition we use a propagation model independent injection rate \cite{Katz:2009yd} given by
\begin{equation}
\begin{split}
Q_{\bar{p},CR}(E)\simeq& 8.4 \times 10^{-33}\text{cm}^{-3}\text{s}^{-1}\text{GeV}^{-1} \left( \frac{E}{100\text{ GeV}}\right)^{-2.8}  \\&\left(1-0.22 \log_{10}^{2}\left(\frac{E}{500\text{ GeV}}\right)\right) \frac{J_{p}(1 \text{ TeV})}{J_{p,0}(1\text{ TeV})}
\end{split}
\end{equation}
where$\ J_{p}(1 \text{ TeV})$\ is the local proton flux at$\ E=1 \text{ TeV}$\ and scaled to measured value$\ J_{p,0}(1 \text{ TeV})\simeq 8\times 10^{-9}\text{ GeV}^{-1}\text{cm}^{-2}\text{s}^{-1}\text{sr}^{-1}.$\ Due to uncertainty in the derivation of the injection rate, it varies within a factor of $2$ \cite{Katz:2009yd}.

The results of our analysis are shown in Fig.~\ref{Ex_total}. Also in this case for the points predicting a too low relic density we assume that our DM candidate is the only source of antiprotons. 

Furthermore, by adopting the Cosmic Rays (CR) grammage given in \cite{Katz:2009yd}, we compute the antiproton flux and compare antiproton to proton flux to measured $\bar{p}/p$ data reported by AMS-02 \cite{Aguilar:2015ooa,Giesen:2015ufa,AMS_Days}.

Fig. \ref{pbarp_tot} presents the allowed region by imposing that the computed $\bar{p}/p$ ratio does not exceed the $\bar{p}/p$ measured by AMS-02. We found that the points reproducing a nearly exact DM relic density do not give significant antiproton flux, and points fitting the $\bar{p}/p$ flux predict a too low relic density. Note that the allowed region can be significantly changed by precise determination of CR grammage and proper knowledge on spallation loss, propagation and solar modulation. In addition, we find that parameter points which generate resonant Sommerfeld enhancement factor are excluded by AMS-02 data. For sake of illustration we provide the $\bar{p}/p$ flux spectra for two points in the Sommerfeld enhanced region in Fig. \ref{pbarp}: data points are the measured $\bar{p}/p$ flux ratio reported by \cite{AMS_Days}, the red line is the secondary prediction as given by \cite{Giesen:2015ufa}, the blue area is the deviation of the secondary prediction due to uncertainities. Model predictions are computed at two parameter points, where $(1)$ is $f = 1500$ GeV, $m_\eta = 1866$ GeV and $m_\sigma = 1303$ GeV, and $(2)$ is $f = 1000$ GeV, $m_\eta = 1183$ GeV and $m_\sigma = 746$ GeV.

\begin{figure*}[!t]
\centering
\includegraphics[width=\columnwidth]{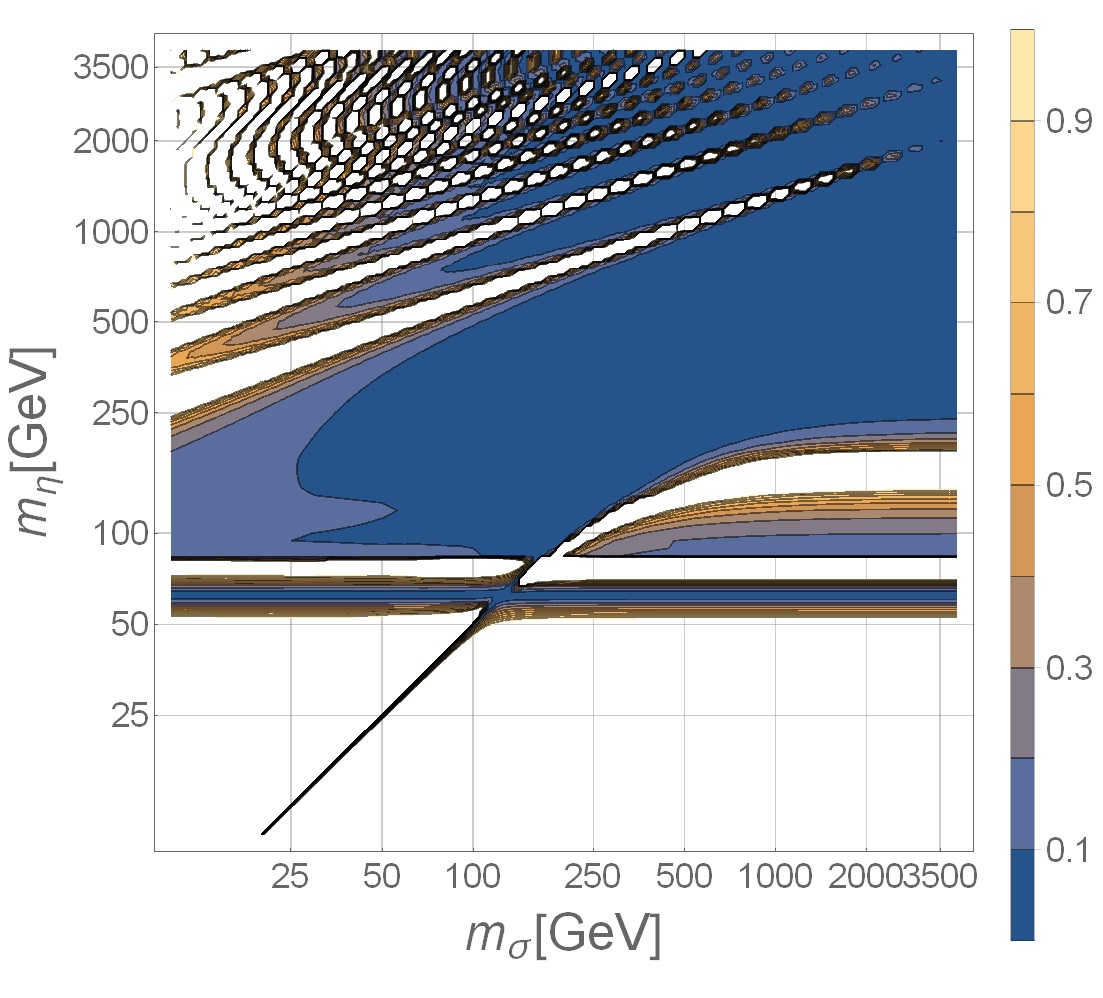}
\includegraphics[width=\columnwidth]{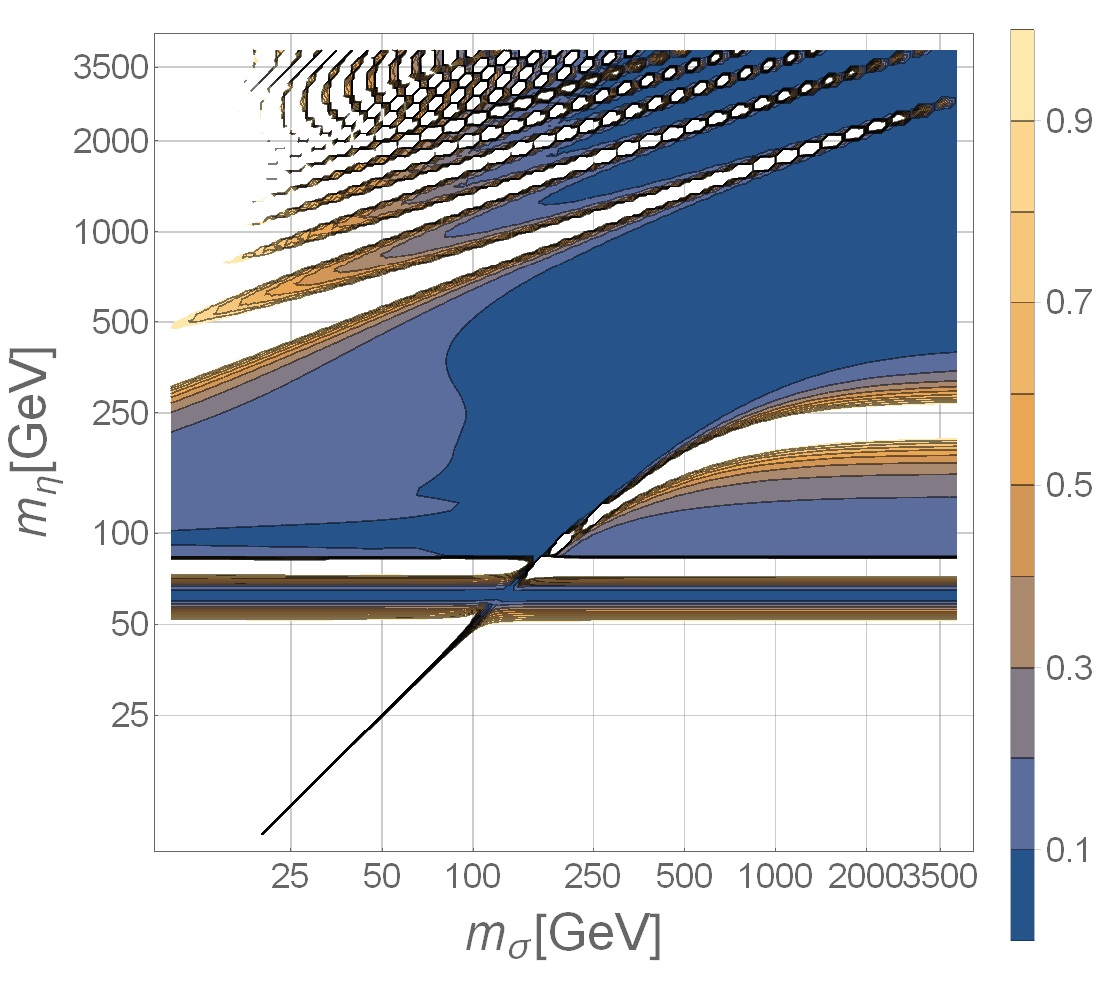}
\caption{Allowed parameter region at $f=1000$ GeV (left) and $f=1500$ GeV (right), comparing with the Fermi-LAT data at $95\%$ confidence level.  Each contour represents a different level for the value of the ratio of ${\sigma_{\bar{b}b} v}$. over the Fermi-LAT bound.}\label{Gamma_tot2}
\end{figure*}

\begin{figure}[t!]
\centering
\includegraphics[width=\columnwidth]{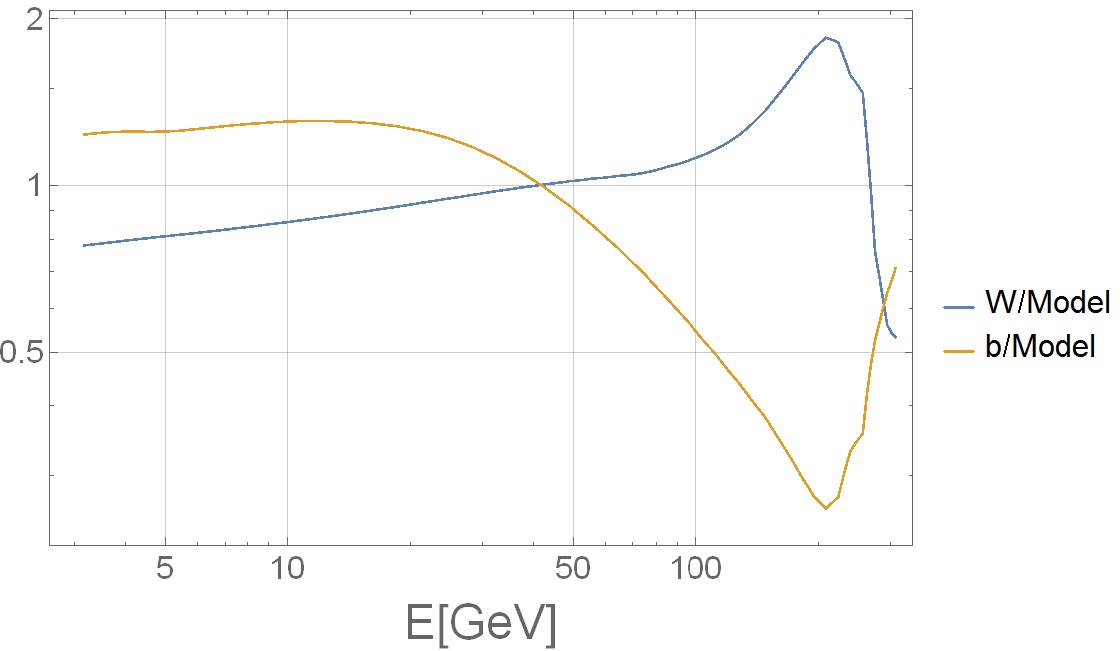}\caption{Gamma ray spectrum ratio for various channels.}\label{Ratio_Spect}
\end{figure}

\subsection{Gamma Ray Flux}\label{Gamma_Ray_Excess}
As well known, see for instance \cite{ArkaniHamed:2008qn}, gamma ray excesses can be a good probe of DM. Since, in our model, DM annihilation produces gamma ray via direct annihilation and Higgs and dilaton mediation, we check whether our model fits the experimental data. Because of the fact that the dwarf spheroidal satellite galaxies (dSphs) of the Milky Way are expected to contain considerable DM amount \cite{McConnachie:2012vd} and have ignorable noise of non-thermal astrophysical gamma ray production, we use the limit on thermally averaged scattering cross sections observed by the Fermi-LAT Collaboration \cite{Ackermann:2015zua} to constrain our model. Note that the analysis is relatively insensitive to the detailed DM distribution inside the dSphs.

\begin{figure}[!h]
\centering
\includegraphics[width=1\columnwidth]{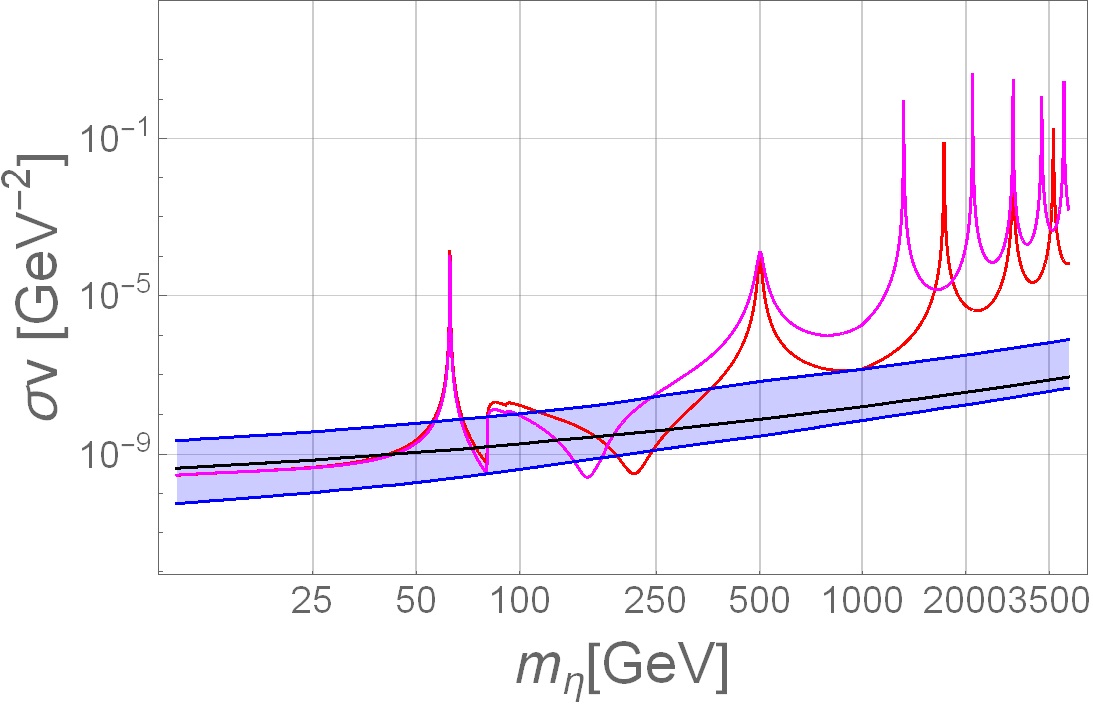}\caption{Thermally averaged cross section in $\bar{b}b$: the magenta curve is computed at $f=1000$ GeV and the red curve at $f=1500$ GeV, fixing $m_\sigma=1000$ GeV. The black line is the constraint for $\bar{b}b$ channel determined by the Fermi-LAT Collaboration, and the blue area is the $2\sigma$ uncertainty.}\label{Gamma_Bound}
\end{figure}
\begin{figure*}[t]
\centering
\includegraphics[width=\columnwidth]{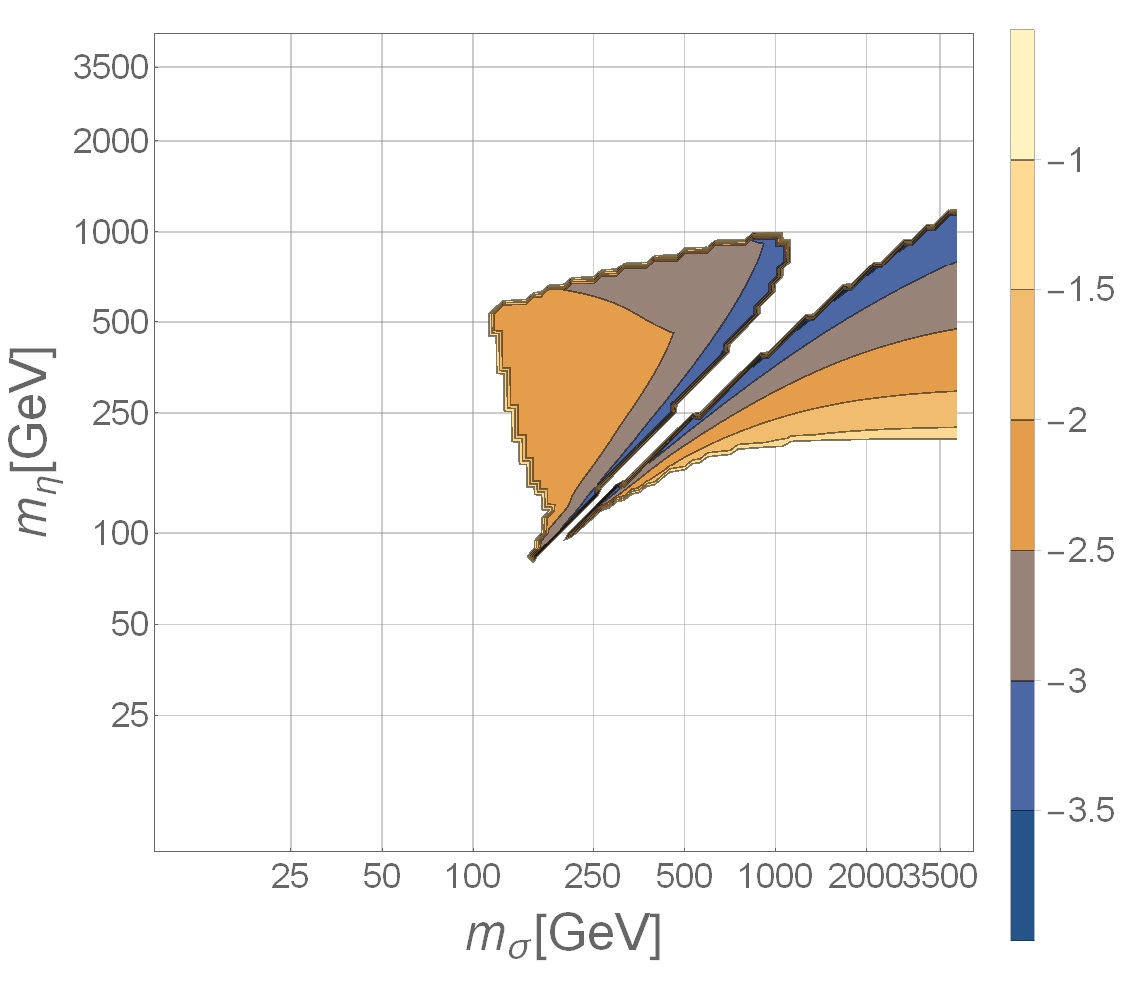}
\includegraphics[width=\columnwidth]{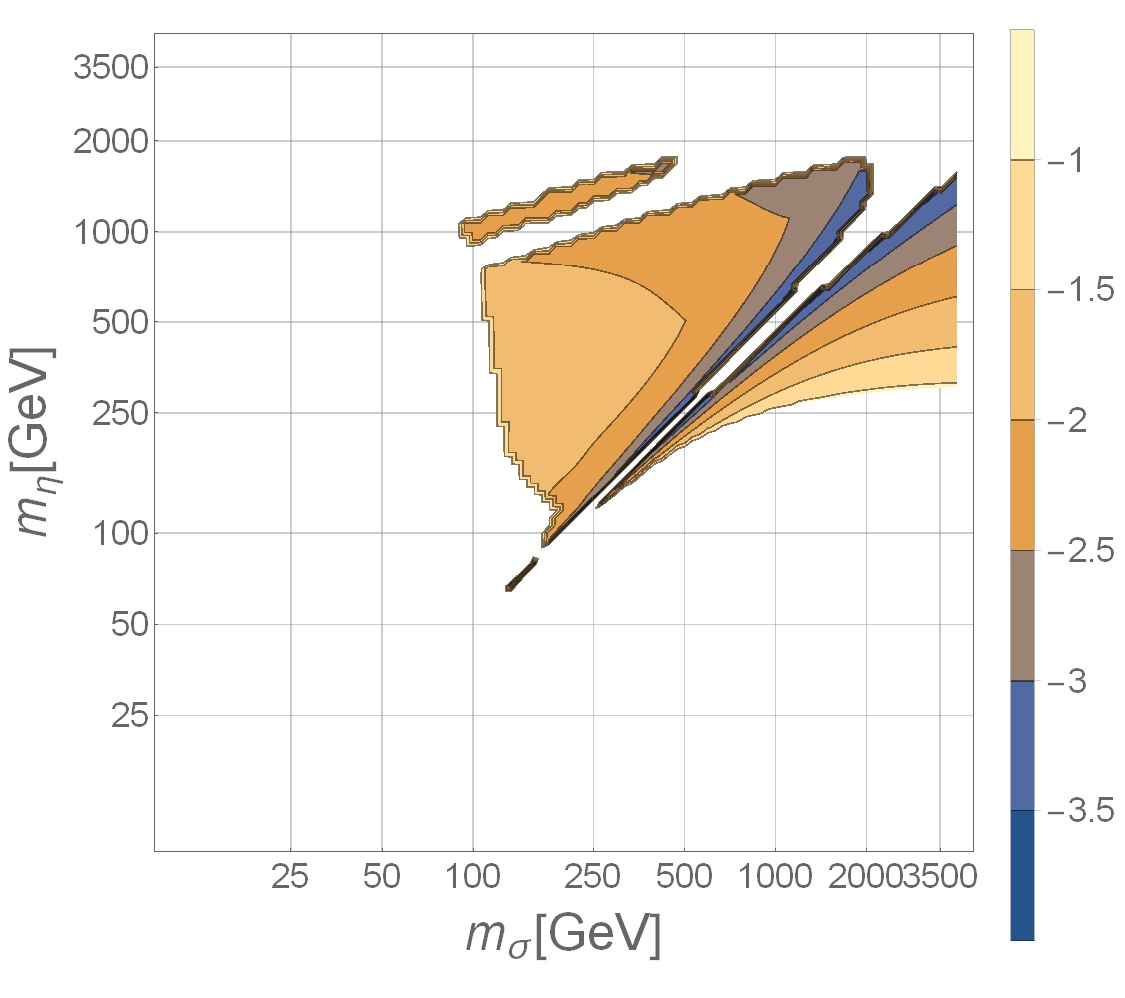}\caption{Relic density of DM, as $\log_{10}(\Omega_\eta h^2)$, fixing $f=1000$ GeV (left) and $ f=1500$ GeV (right) taking into account all the constraints discussed in the text.}\label{Sum_1500}
\end{figure*}

Following \cite{Cirelli:2010xx,Elor:2015tva} we compute the gamma ray spectrum per annihilation, and we compare with the SM channels, which we find in \cite{Cirelli:2010xx}. Fig. \ref{Ratio_Spect} shows the ratio of gamma ray spectrum at each energy. The spectrum generated by DM annihilation of our model is within a factor of $2$ or $3$ with respect to the gamma ray spectrum generated by pure $\eta \eta\rightarrow \bar{b}b$ channel and $\eta \eta \rightarrow WW$ channel, thus we assume that the constraints given by \cite{Ackermann:2015zua} is applicable to our model.

In many points of the parameter space the correct relic density of DM is not reproduced, as we discussed above and we showed in Fig.~\ref{Relic2_total}. For those points we assume that $\eta$ only partially accounts for the DM density around the dSphs and the additional DM does not contribute to the CR production.

Under such assumptions the resulting effective J factor contributing to the gamma ray flux is
\begin{equation}
J_{eff}=(\frac{\Omega_{\eta}}{\Omega_{DM}})^{2} J
\end{equation}
where $\Omega_{DM}h^{2} \simeq 0.12$ and $\Omega_{\eta}$ is the relic density for $\eta$ DM. Consequently we derive a cross section bound much weaker the bound given by \cite{Ackermann:2015zua}.

Fixing $m_\sigma =1000$ GeV we present thermally averaged cross section and bounds given by the Fermi-LAT Collaboration in Fig.~\ref{Gamma_Bound}. 
Fig.~\ref{Gamma_tot2} shows the allowed parameter region imposing the constraints from the Fermi-LAT experiment at $95\%$ confidence level. We do not observe any peak in the gamma ray spectrum because $\sigma v_{\eta\eta\rightarrow \gamma\gamma}/\sigma v_{tot}$ is negligible in our model. 

In the high DM mass region, where$\ m_\eta \geq1\text{ TeV},$\ experimental constraints given by the H.E.S.S Collaboration \cite{HESS:2015cda} provide tighter bound though we have more dependence on the propagation model. By assuming that DM distribution follows a cusp distribution such as  the Navarro-Frenk-White \cite{Navarro:1995iw}, we could superimpose this additional bound on the constraints given by Fermi-LAT, but that region is already ruled out and this procedure does not provide additional information.

\section{Collider Constraints}\label{sec: collider}

\subsection{Higgs Measurements}

\begin{figure}[t]
\centering
	\includegraphics[width=\columnwidth]{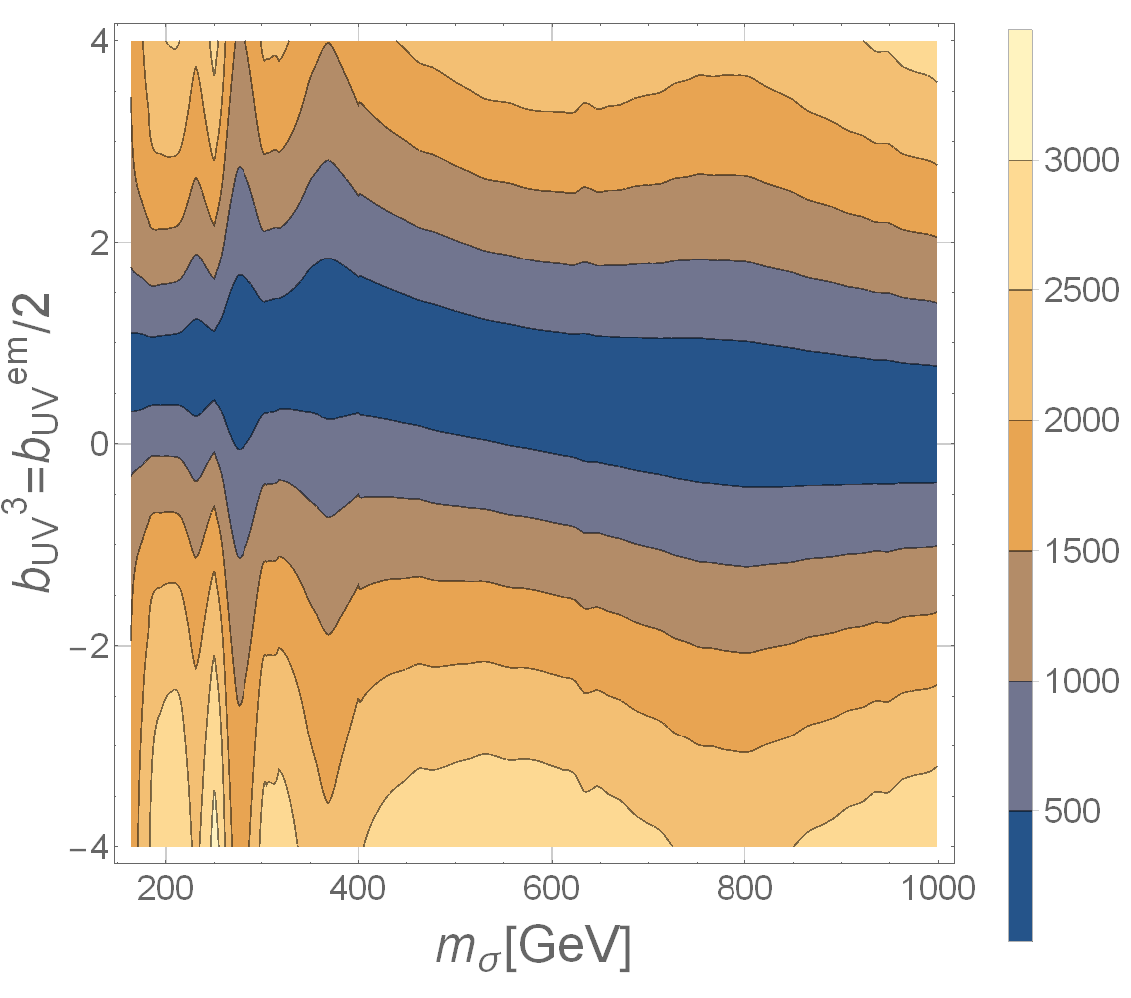}\caption{95\% CL lower bound on the symmetry breaking scale $f$ in GeV varying the dilaton mass and the UV beta functions from searches for heavy scalars.}\label{Heavy_Scalar_jpg}
\end{figure}

We consider the impact of the measurements of signal strengths reported in \cite{ATLAS:2015kwa,CMS:2015kwa} on the allowed parameter space of the theory, namely on $\xi$ or equivalently on $f$. We perform a $\chi^2$ analysis using the following channels
\begin{equation}
\begin{split}
\mu_V /\mu_F =1.06^{+0.35}_{-0.27}\,,\quad \mu_{F}^{\gamma\gamma}=1.13^{+0.24}_{-0.21}\,,\\
\mu_{F}^{ZZ}=1.29^{+0.29}_{-0.25}\,,\quad\mu_{F}^{WW}=1.08^{+0.22}_{-0.19}\,,\\
\mu_{F}^{\tau\tau}=1.07^{+0.35}_{-0.28}\,,\quad\mu_{F}^{bb}=0.65^{+0.37}_{-0.28}.
\end{split}
\end{equation}
and the result is shown in Fig.~\ref{Collider_jpg}, from which we read that at 95$\%$ CL $f$ larger than $960$ GeV is still allowed.

\subsection{Heavy Scalar Searches}

Since the dilaton has couplings to SM particles similar to the Higgs' ones its parameter space is constrained by searches for heavy Higgses \cite{TheATLAScollaboration:2013zha,Chatrchyan:2013yoa,Khachatryan:2015cwa}. A dilaton whose mass lies between $200$ and $1000$ GeV is probed by such searches, and the experimental measures convert to a lower bound on $f$. In Fig.~\ref{Heavy_Scalar_jpg} we report the allowed minimum value for $f$ at $95\%$ CL for each choice of dilaton mass, focusing for definiteness on specific values for the UV beta functions $b_{UV}^{3,em}$, chosen as representative.

\subsection{Precision Tests}
We proceed inspecting the contribution of new physics to the EW precision parameters measured by LEP \cite{LEP:2003aa}. The presence of composite resonances is expected to have an impact on EW precision tests .At tree level vector resonances give, imposing the generalized Weinberg sum rules as in  \cite{Marzocca:2012zn},
\begin{align}
\delta S &=\frac{8 \sin^{2}\theta_w m^{2}_{W}}{\alpha m^{2}_{\rho}}\left(1-\frac{f^2}{4 f^{2}_{\rho}}\right)
\end{align}
which in turn implies for instance $m_\rho>2$ TeV if $f_\rho=f$.
Also modification of Higgs couplings play a role in enhancing EW precision parameters: interestingly enough once we include the dilaton we get vanishing $T$ corrections due to the fact that $ c_{V,h}^2 +c_{V,\sigma}^{2} =1$. Furthermore, for the same reason, $S$ correction are also suppressed. Higgs and dilaton loops are computed following \cite{Falkowski:2013dza}. From the Lagrangian
\begin{align}
\mathcal{L}\supseteq& \left(2m^{2}_{W} W^{+}_\mu W^{- \mu}+m^{2}_{Z} Z_\mu Z^\mu\right) \left( c_{V,h}\frac{h}{v}+c_{V,\sigma}\frac{\sigma}{v}  \right)\nonumber\\
& -\frac{1}{4}\frac{\sigma}{v}(2c_{Z\gamma}F_{\mu\nu}Z^{\mu\nu}+c_{\gamma\gamma}F_{\mu\nu}F^{\mu\nu})
\end{align}
we easily read
\begin{align}
\alpha\Delta T \simeq& -\frac{3 g^{2}_{Y}}{32\pi^{2}}(1-c^{2}_{V,h}-c^{2}_{V,\sigma})\log(\Lambda/m_Z)=0\,,\nonumber\\
\alpha\Delta S \simeq& \frac{g_L g_Y\log(\Lambda/m_Z)}{48 \pi^{2}(g_{L}^{2}+g_{Y}^{2})}(2 g_L g_Y (1-c^{2}_{V,h}-c^{2}_{V,\sigma})\nonumber\\
&+6c_{V,\sigma}(2 g_L g_Y c_{\gamma\gamma}+c_{Z\gamma}(g^{2}_L -g^{2}_Y))\nonumber\\
&+3(g_L g_Y(c^{2}_{Z,\gamma}-c^{2}_{\gamma\gamma})-(g_{L}^{2}-g^{2}_{Y})c_{\gamma\gamma}c_{Z\gamma}))\,,\nonumber\\
\alpha\Delta W\simeq& \frac{g^{2}_{L}}{192\pi^{2}}\left(c_{\gamma\gamma}+\frac{g_L}{g_Y}c_{Z\gamma}\right)^{2}\log(\Lambda/m_Z)\,,\nonumber\\
\alpha\Delta Y\simeq &\frac{g^{2}_{L}}{192\pi^{2}}\left(c_{\gamma\gamma}-\frac{g_L}{g_Y}c_{Z\gamma}\right)^{2}\log(\Lambda/m_Z)\,,
\end{align}
where $\Lambda\simeq4\pi f$ and in our model 
\begin{align} 
c_{V,h}=&\sqrt{1-\xi}\,,\quad c_{V,\sigma}=\sqrt{\xi}\,,\\
c_{\gamma\gamma}=& -\frac{\alpha_{em}}{2\pi}(b_{IR}^{em}-b_{UV}^{em}+\frac{4}{3}F_{1/2}(x_t)-F_{1}(x_W))\sqrt{\xi}\,,\nonumber\\
 c_{Z,\gamma} =&-\frac{\alpha_{em}}{2\pi \tan \theta_W}(b_{IR}^{2}-b_{UV}^{2}-t_{W}^{2} (b_{IR}^{1}-b_{UV}^{1}))\sqrt{\xi}\nonumber\\
 &+\frac{e g_L}{8 \pi^{2}}(A_{1}^{Z}(\tau_W,\lambda_W)+\sum_{f}N_{f}q_f g_f A^{Z}_{1/2}(\tau_f,\lambda_f)) \sqrt{\xi}\,,\nonumber
 \end{align}
 with $A_1$ and $A_{1/2}$ given in \cite{Korchin:2013ifa}.

As a result EW precision tests do not significantly constraint the model for $f\geq900$ GeV. Finally note that typical values of $\alpha W$ and $\alpha Y$ are $\sim 10^{-7}$. Fermionic resonances are expected to affect EW parameters as well but in a model dependent way: we rely on the fact that this effect is well studied and understood in the literature and it is shown to be compatible with observations for large regions in parameters space in similar models.

\section{Summary and Conclusions}\label{sec: summary}

The presence of additional light scalars, beyond the Higgs, is an expected feature of CHM. We have considered a candidate DM scalar particle in a specific CHM based on the coset $\SO(6)/\SO(5)$, enlightening the possible role of a light dilaton as a mediator of DM interactions with the SM. To summarize our analysis we combine results from collider constraints, direct and indirect searches discussed in the previous sections. Fig.~\ref{Sum_1500} shows the predicted density for two given symetry breaking scales $f=1000$ GeV and $f=1500$ GeV. For these plots we use benchmark UV beta functions $b^{3}_{UV}=b^{em}_{UV}=0$. While for $f=1000$ GeV the available parameter space, in which our candidate DM scalar entirely accounts for the observed density, shrinks to zero, if we allow for $f=1500$   GeV we have a region in parameter space starting with $m_\eta\simeq200$~GeV and $m_\sigma\simeq500$~GeV; a heavier dilaton requires a heavier DM particle and an asymptotic value of $m_\eta\simeq300$ GeV is reached at $m_\sigma\simeq1500$ GeV. Interestingly, according to the scan performed in \cite{Marzocca:2014msa}, $\eta$ mass can vary between $100$ and $700$ GeV for $f=800-1100$ GeV. Notice that $f=1000$ GeV returns to be a viable option if a fraction of the DM relic density is accounted for by a different particle, as for instance an axion.

\section*{Acknowledgments}

M.K. thanks KAIST Undergraduate Research Program (URP) for its support, with which this project was initiated. 
S.L. has been supported in part by the National Research Foundation of Korea(NRF) grant funded by the Korea government(MEST) (NRF-2015R1A2A1A15052408).

\appendix

\begin{figure}[!t]
\centering
	\includegraphics[width=\columnwidth]{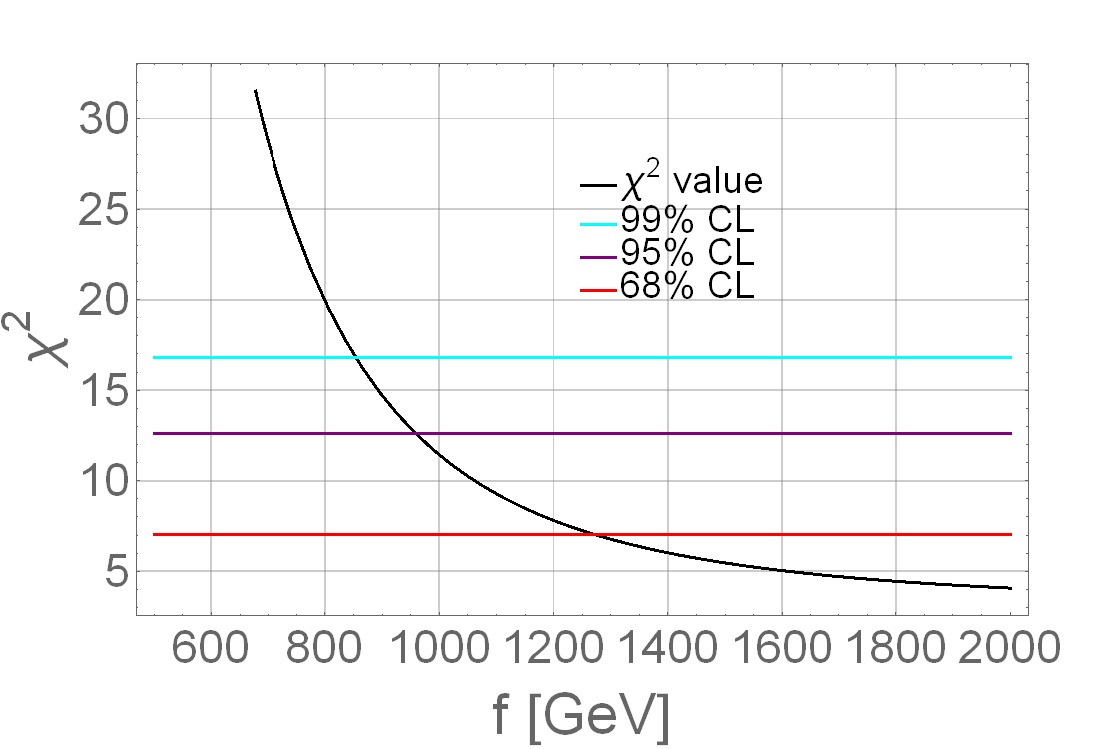}\caption{$\chi^{2}$ value varying $f$ from the discussed Higgs channels.}\label{Collider_jpg}
\end{figure}

\section{Details of the Models}\label{app: details}
\subsection{Fermionic Sector}
The Lagrangian of the fermionic sector, including composite resonances, is given by
\begin{align}\label{eq:top pc}
\noindent \mathcal{L}_{f}=&\bar{q}_Li \slashed{D}q_L +\bar{t}_{R} i\slashed{D}t_R \\
&+\sum_{i}\bar{S}_{i}(i\slashed{\nabla}-m_{iS})S_{i} + \sum_{j} \bar{F}_{j}(i\slashed{\nabla}-m_{jF})F_{j}\nonumber\\
&+ \sum_{i}(\epsilon^{i}_{tS}\bar{\xi}_{R}P_{L}US_{i} + \epsilon^{i}_{qS}\bar{\xi}_{L}P_{R}US_{i})+h.c\nonumber\\
&+ \sum_{j}(\epsilon^{j}_{tF}\bar{\xi}_{R}P_{L}UF_{j}+\epsilon^{j}_{qF}\bar{\xi}_{L}P_{R}UF_{j})+h.c\,.\nonumber
\end{align}
In addition, there can be interactions between composite resonances \cite{Marzocca:2012zn,Marzocca:2014msa}:
\begin{align}
\mathcal{L}_{int}=&\sum_{\eta=L,R}( k^{V, \eta}_{ij}\bar{F}_{i}\gamma^{\mu}(g_\rho \rho_\mu -E_\mu ) P_\eta F_j) \nonumber\\
&+\sum_{\eta=L,R}(\bar{S}_{i}\gamma^{\mu}(k^{A,\eta}_{ij}a_{\mu}+k^{d,\eta}_{ij}d_\mu)P_\eta F_j +h.c)\,.\nonumber
\end{align}
where $\rho_\mu$, $a_\mu$ are massive vector resonances of the strong sector. 
Notice that these interactions do not enter the scalar couplings to $gg$ and $\gamma\gamma$ at one loop because they mix different species of composite fermions \cite{Montull:2013mla}.

In order to compute the low energy effective theory of SM fermions, we need to integrate out the composite resonances. The result, in momentum basis, up to quadratic order in the fermions, is written as
\begin{equation}
\mathcal{L}_{eff}=\Pi_{t_L}\bar{t}_L \slashed{p} t_L+\Pi_{t_{R}}\bar{t}_{R}\slashed{p}t_{R}-(\Pi_{t_{L}t_{R}}\bar{t}_{L}t_{R}+h.c)
\end{equation}
The form factors are written as
\begin{align}
\Pi_{t_L}=&\Pi_F +\frac{h^2}{f^2}\Pi_{1F}\,,\,\,
\Pi_{t_R}=\Pi_S +(1-\frac{h^2}{f^2}-\frac{\eta^2}{f^2})\Pi_{1S}\,,\nonumber\\
\Pi_{t_L t_R}=&\frac{h}{f}\sqrt{1-\frac{h^2}{f^2}-\frac{\eta^2}{f^2}}\Pi_{FS}\,.
\end{align}
The explicit form of the form factors in terms of the parameters in \refeq{eq:top pc} is given in \cite{Marzocca:2014msa}.

\subsection{Vector Resonances}
The Lagrangian for vector resonances is given by
\begin{align}\label{eq: vector pc}
\mathcal{L} =&-\frac{1}{4}Tr(\rho^{2}_{\mu \nu})+\frac{f_{\rho}^{2}}{2}Tr(g_\rho \rho_\mu -E_\mu)^2\nonumber\\
&-\frac{1}{4}Tr(a^{2}_{\mu \nu})+\frac{f_{a}^{2}}{2 \Delta^2}Tr(g_a a_\mu -\Delta d_\mu )^2\,.
\end{align}

General cases of vector resonances are examined in \cite{Marzocca:2012zn} and mixing between $\rho$ and $E$ is described in \cite{Marzocca:2012zn,Marzocca:2014msa}.
Similarly to the fermion case, integrating out heavy vector fields we obtain an effective Lagrangian for SM vector bosons given by, in momentum space, 
\begin{align}
\mathcal{L}=&\frac{P^{\mu\nu}_T}{2}(\Pi_{0}(q^2)\Tr(A_\mu A_\nu) + \Pi_{1}(q^2)\Sigma^t A_\mu A_\nu \Sigma\nonumber\\
&+\Pi_{0}^{X}(q^2)X_\mu X_\nu)
\end{align}
where $A_\mu$ is a spurion obtained formally gauging all the $\SO(5)$ generators. In the physical configuration where only $A^{a_L}=W^{a} $, $A^{3_R}=c_X B$, $X=s_X B$ are different from zero, with $c_X=g_{Y}/g_L$ and $s^{2}_{X}=1-c^{2}_{X}$, the former expression reduces to
\begin{align}
\mathcal{L}=&\frac{P^{\mu \nu}_{T}}{2}( \Pi_0 W_{\mu}^a W_{\nu}^a +\Pi_1 \frac{h^2}{4 f^2}(W^{1}_\mu W^{1}_\nu +W^{2}_\mu W^{2}_\nu)\nonumber\\
&+ \Pi_B B_\mu B_\nu + \Pi_1 \frac{h^2}{4 f^2 \cos^{2}\theta_w}Z_\mu Z_\nu )
\end{align}
where $\Pi_B=s^{2}_{X}\Pi^{X}_0 +c^{2}_{X}\Pi_0$ and $Z=\cos\theta_w W-\sin\theta_w B$, with $\cos\theta_w = g_L /\sqrt{g^{2}_L +g^{2}_Y}$. It is also customary to define
\begin{align}
\Pi_{WW}=&\Pi_0 +\frac{h^2}{4 f^2}\Pi_1\,,\quad
\Pi_{BB}=\Pi_B +c^{2}_{X}\frac{h^2}{4 f^2} \Pi_1\,,\nonumber\\
\Pi_{W_3 B} =&-c_{X} \frac{h^2}{4 f^2}\Pi_1\,.
\end{align}
 
 \subsection{Dilaton Potential}\label{app: dilaton}
Unlike other Goldstone bosons, a non derivative self-interaction term for the dilaton is allowed and indeed it is expected at tree level:
\begin{equation}
V_{tree}(\chi)= \frac{\kappa}{4!} \chi^4\,.
\end{equation}
Corrections are generated by loops of self interactions and loops of heavy resonances. The first gives\begin{align}
V_{eff}=&V_{tree}+\frac{3 \kappa^2}{32 \pi^2} \chi^4 \left( \log \frac{\kappa \chi^{2}}{2 \mu^2} -\frac{1}{2}\right)\nonumber\\
=& \frac{1}{32\pi^2}\frac{\chi^4}{f^4}\left( \hat{\kappa}_0 \log \frac{\chi^2}{f^2}+\hat{\kappa}_1 \right)
\end{align}
where 
\begin{align}
\hat{\kappa}_0 = &3\kappa^{2} f^{4}\,, \nonumber\\
\hat{\kappa}_1 = &\frac{32 \pi^2 \kappa f^{4}}{4!}+3\kappa^2 f^4\left(\log\frac{\kappa f^{2}}{2\mu^{2}}-\frac{1}{2}\right) \,.
\end{align}
Gauge and fermion contributions to the potential are obtained from the form factors at $h=\eta=0$:
\begin{align}
V(\chi)&=\int\frac{d^4 p_{E}}{{(2 \pi )}^4}\left(\frac{3}{2} \log[ \Pi_{0}\Pi_{B}]\right.\nonumber\\
&-6\log[p^{2}_{E}\Pi_{F}(\Pi_{1S}+\Pi_{S})]\Big)\,.
\end{align}
Recalling the general formula
\begin{equation}
\int\frac{d^{4} p_{E}}{(2\pi)^{4}}\log[p^{2}_{E}+U^2]=\frac{1}{32 \pi^{2}}U^{4}(\log\frac{U^2}{\mu^{2}}-\frac{1}{2})
\end{equation}
the result can be expressed as
\begin{equation}
V(\chi)\simeq\frac{1}{32\pi^{2}}\frac{\chi^4}{f^4}(\kappa_0 \log \frac{\chi^2}{f^2}+\kappa_1)
\end{equation}
where trivially
\begin{align}
\kappa_0=\frac{2\pi^2f^4\chi}{3}\frac{\partial^5 V}{{\partial\chi}^5}\,,\,\kappa_1=\frac{4\pi^2f^4}{3}{\left.\frac{\partial^4 V}{{\partial\chi}^4}\right|}_{\chi\rightarrow f}-\kappa_0\,.
\end{align}

We now move to study the Vacuum Expectation Value (VEV) and the mass of the dilaton. We start with the potential  
\begin{equation}
V_{eff}(h,\eta,\chi)=\frac{\chi^4}{f^4} V(h,\eta)+\frac{1}{32\pi^2}\frac{\chi^{4}}{f^4}(\kappa_0 \log\frac{{\chi^2}}{f^2}+\kappa_1)
\end{equation}
where $V(h,\eta)$ is the sum of the gauge and fermion contributions. 
Imposing the condition $\langle\chi\rangle=f$ we obtain
\begin{equation}
\kappa_1 = -32 \pi^{2} V(v,0)-\frac{\kappa_0}{2}\,,
\end{equation}
and then
\begin{align}\label{eq:dilaton pot}
V_{eff}(h,\eta,\chi)=&\frac{\chi^4}{f^4}(V(h,\eta)-V(v,0))\\
&+\frac{\kappa_0}{16\pi^2}\frac{\chi^{4}}{f^4}( \log\frac{\chi}{f}-\frac{1}{4})\,.\nonumber
\end{align}
Therefore the mass of the dilaton is given by
\begin{equation}
m^{2}_{\sigma}=\frac{\kappa_0}{4\pi^{2}f^{2}}
\end{equation}
and the effective potential \refeq{eq:dilaton pot} can be rewritten as  
\begin{equation}\label{eq:Veff dilaton}
V_{eff}(h,\eta,\chi)=\frac{\chi^4}{f^4}(V(h,\eta)-V(v,0))+\frac{m^{2}_{\sigma}}{4}\frac{\chi^{4}}{f^2}(\log\frac{\chi}{f}-\frac{1}{4})\,.
\end{equation}
We assumed $\kappa_0>0$ in order to have a potential bounded from below. Finally we notice that because of the tree level term the dilaton mass is model dependent and therefore in our phenomenological analysis we treat it as a free parameter.

\section{Decoupling of Heavy Composite Fermions}\label{app: decoupling}
We discuss here the effect of heavy fermionic resonances on the couplings of the dilaton $\sigma$ to $\gamma\gamma$ and $gg$. They contribute entering the beta function coefficients $b_{IR}^i$ and also circulating in triangular loops. In the limit of mass much larger than $m_\sigma/2$ the two effects cancel and in the following we review this property. Indeed in extra dimensional construction heavy KK modes of bulk fermions do not generate corrections for radion couplings, as shown in \cite{Csaki:2007ns}. We obtain the same result in a four dimensional language.

We consider $N_F$ and $N_S$ heavy Dirac fermions with quantum numbers under the SM gauge group $\SU(N_c)\times{\SU(2)}_L \times{\U(1)}_Y$ 
\begin{equation}
F={(N_c,2)}_{7/6}\oplus{(N_c,2)}_{1/6}\oplus{(N_c,1)}_{2/3},S={(N_c,1)}_{2/3}\,.
\end{equation}
They enter the Lagrangian
\begin{align}
\mathcal{L}\supseteq& \frac{\alpha_s}{8\pi} (b^{3}_{IR,F}+b^{3}_{IR,S})\frac{\sigma}{f} G^{a}_{\mu\nu}G^{a\mu\nu}\nonumber\\
&+\frac{\alpha_{em}}{8\pi}(b^{em}_{IR,F}+b^{em}_{IR,S})\frac{\sigma}{f}F_{\mu\nu}F^{\mu\nu}
\end{align}
contributing with
\begin{align}
b_{IR,F}^{3}=&-\frac{10}{3} N_F\,,\, b^{em}_{IR,F}=\frac{N_c}{27}152N_F  \,, \nonumber\\
b_{IR,S}^{3}=&-\frac{2}{3}N_S\,,\, b_{IR,S}^{em}=\frac{N_c}{27}16N_S\,.
\end{align}
The second contribution comes from loop diagrams. For $\sigma gg$ it has the form
\begin{align}
\mathcal{L}_{eff}&\supseteq\frac{\alpha_s}{8\pi}(\frac{5N_F}{2}F_{1/2}(x_F^2)+\frac{N_S}{2}F_{1/2}(x_S^2))\frac{\sigma}{f}G^{a}_{\mu\nu}G^{a\mu\nu}
\\&=\frac{\alpha_s}{8\pi}(\frac{5N_F}{2}F_{1/2}(x_F^2)+\frac{N_S}{2}F_{1/2}(x_S^2))\frac{\sigma}{f}G^{a}_{\mu\nu}G^{a\mu\nu}\nonumber
\end{align}
where $x_{F,S}=2m_{F,S}/m_\sigma$. Note that $F_{1/2}(x)$ quickly saturates to $4/3$ for $x>1$. Since typical masses of heavy composite fermions are larger than $m_{\sigma}/2$ the limit is justified and we have a perfect cancellation in the infinite mass limit.
Similarly for $\sigma\gamma\gamma$
\begin{align}
\mathcal{L}_{eff}&\supseteq\frac{\alpha_{em}}{8\pi}N_c(\frac{38 N_F}{9}F_{1/2}(x_F^2)+\frac{4N_S}{9}F_{1/2}(x_S^2))\frac{\sigma}{f}F_{\mu\nu}F^{\mu\nu}
\\&=\frac{\alpha_{em}}{8\pi}N_c(N_F\frac{38}{9}F_{1/2}(x_F^2)+N_S\frac{4}{9}F_{1/2}(x_S^2))\frac{\sigma}{f}F_{\mu\nu}F^{\mu\nu}\nonumber
\end{align}
and the same cancellation is in place.
Therefore we verify, at one loop, the decoupling of heavy fermions states, confirming the expectation from extra dimensional models.

\section{Dilaton Decay Widths}\label{app:widths}
\begin{align}
\Gamma_{\sigma \rightarrow \bar{\psi}\psi} =& \frac{3 m^{2}_{\psi} (m^{2}_{\sigma}-4m^{2}_{\psi})^{3/2}}{8\pi f^{2} m^{2}_{\sigma}}\,,\nonumber\\
\Gamma_{\sigma \rightarrow hh}=&\frac{\sqrt{m^{2}_{\sigma}-4m^{2}_{h}} (m^{2}_{\sigma}+2m^{2}_{h})^{2}}{32 \pi f^{2}m^{2}_{\sigma}}\,,\nonumber\\
\Gamma_{\sigma\rightarrow WW}=&\frac{\sqrt{m^{2}_{\sigma}-4m^{2}_{W}}(m^{4}_{\sigma}-4m^{2}_{\sigma}m^{2}_{W}+12m^{4}_{W})}{16 \pi f^{2} m^{2}_{\sigma}}\,,\nonumber\\
\Gamma_{\sigma\rightarrow gg}=&\frac{\alpha_s^2}{32\pi^3}{(b_{IR}^{3}-b_{UV}^{3}+\frac{1}{2}F_{1/2}(x_t))}^2\frac{m_\sigma^3}{f^2}\,,\nonumber\\
\Gamma_{\sigma\rightarrow\gamma\gamma}=&\frac{\alpha^2}{256\pi^3}{(b_{IR}^{em}-b_{UV}^{em}+\frac{4}{3}F_{1/2}(x_t)-F_1 (x_W))}^2 \frac{m_\sigma^3}{f^2}\,,\nonumber\\
\Gamma_{\sigma\rightarrow\eta\eta}=&\frac{\sqrt{m^{2}_{\sigma}-4m^{2}_{\eta}} (m^{2}_{\sigma}+2m^{2}_{\eta})^{2}}{32 \pi f^{2}m^{2}_{\sigma}}\,.
\end{align}

\section{Annihilation Cross Sections}\label{appendix:annihilation}
\begin{widetext}
\begin{align}
\sigma v_{\eta \eta\rightarrow WW}=&\frac{m_{W}^{4} \sqrt{m^{2}_{\eta}-m^{2}_{W}}}{32 \pi m_{\eta}^{3} f^4}\left(2+\left(\frac{2 m^{2}_{\eta}-m^{2}_{W}}{m^{2}_{W}}\right)^2\right)\left[ \frac{144  m^{4}_{\eta}}{|m^{2}_{\sigma}-4m^{2}_{\eta}+i \Im(\Pi_{\sigma}(4m^{2}_{\eta}))|^2}  \right.\nonumber\\
&+ 48 m^{2}_{\eta} m^{2}_{W} \frac{(-4 m^{2}_{\eta} +f^{2} \lambda_{h \eta}(1-\xi))(16m^{4}_{\eta}-4 m^{2}_{\eta}(m^{2}_{\sigma}+m^{2}_{h})+m^{2}_{h}m^{2}_{\sigma}+\Im(\Pi_{\sigma}(4m^{2}_{\eta}))\Im(\Pi_{h}(4m^{2}_{\eta}))}{|m^{2}_{\sigma}-4m^{2}_{\eta}+i \Im(\Pi_{\sigma}(4m^{2}_{\eta}))|^2 |m^{2}_{h}-4m^{2}_{\eta}+i \Im(\Pi_{h}(4m^{2}_{\eta}))|^2}\nonumber\\
&\left.+4 \frac{(4 m^{2}_{\eta}-f^{2}\lambda_{h \eta} (1-\xi))^2}{|m^{2}_{h}-4m^{2}_{\eta}+i \Im(\Pi_{h}(4m^{2}_{\eta}))|^2}\right]\,,
\end{align}
\begin{align}
\sigma v_{\eta \eta \rightarrow \sigma\sigma}=&\frac{\sqrt{m^{2}_{\eta}-m^{2}_{\sigma}
}}{4\pi f^{4} (2m^{2}_{\eta}-m^{2}_{\sigma})^2 |m^{2}_{\sigma}-4m^{2}_{\eta}+i \Im(\Pi_{\sigma}(4m^{2}_{\eta}))|^2 }(49 m^{8}_{\sigma} m_{\eta}-108 m^{4}_{\sigma}m^{5}_{\eta}+32 m^{2}_{\sigma}m^{7}_{\eta}+64m^{9}_{\eta}-28m^{6}_{\sigma}m^{3}_{\eta})\nonumber\\
&+\frac{\sqrt{m^{2}_{\eta}-m^{2}_{\sigma}
}}{4\pi f^{4} (2m^{2}_{\eta}-m^{2}_{\sigma})^2 |m^{2}_{\sigma}-4m^{2}_{\eta}+i \Im(\Pi_{\sigma}(4m^{2}_{\eta}))|^2 }(25m^{4}_{\sigma}m_{\eta}+4m^{5}_{\eta}-20m^{2}_{\sigma}m^{3}_{\eta})(\Im(\Pi_{\sigma}(4m^{2}_{\eta})))^2 )\,,
\end{align}
\begin{align}
\sigma v_{\eta \eta\rightarrow\bar{\psi}\psi}=&\frac{3 m^{2}_{\psi}}{8 \pi m^{3}_{\eta}}{(m^{2}_{\eta}-m^{2}_{\psi})}^{3/2}\left[\frac{36 m^{4}_{\eta}}{f^{4} |m^{2}_{\sigma}-4m^{2}_{\eta}+i \Im(\Pi_{\sigma}(4m^{2}_{\eta}))|^2} +\frac{16 (4m^{2}_{\eta}-f^{2} \lambda_{h \eta}(1-\xi)^{3/2})^{2} (1-2\xi)}{f^{4} |m^{2}_{h}-4m^{2}_{\eta}+i \Im(\Pi_{h}(4m^{2}_{\eta}))|^2 (1-\xi)^{3}}\right.\nonumber\\
& -\frac{48(-4m^{2}_{\eta} m^{2}_{\sigma} +16m^{4}_{\eta}+m^{2}_{\sigma}m^{2}_{h}-4m^{2}_{\eta}m^{2}_{h}+\Im(\Pi_{\sigma}(4m^{2}_{\eta}))\Im(\Pi_{h}(4m^{2}_{\eta})))(4m^{2}_{\eta}-f^{2} \lambda_{h \eta}(1-\xi)^{3/2})(1-2\xi)}{|m^{2}_{h}-4m^{2}_{\eta}+i \Im(\Pi_{h}(4m^{2}_{\eta}))|^2 |m^{2}_{\sigma}-4m^{2}_{\eta}+i \Im(\Pi_{\sigma}(4m^{2}_{\eta}))|^2 (1-\xi)^{3/2}}\nonumber\\
&+\frac{8 (4m^{2}_{\eta}-m^{2}_{h})(4m^{2}_{\eta}-f^{2}\lambda_{h \eta}(1-\xi)^{3/2})(1-2\xi)\xi}{v^{2} f^{2}(1-\xi)^{5/2} |m^{2}_{h}-4m^{2}_{\eta}+i \Im(\Pi_{h}(4m^{2}_{\eta}))|^2}+\nonumber\\
&\left.+\frac{12m^{2}_{\eta}(m^{2}_{\sigma}-4 m^{2}_{\eta})\xi}{f^{2} v^{2} (1-\xi)|m^{2}_{\sigma}-4m^{2}_{\eta}+i \Im(\Pi_{\sigma}(4m^{2}_{\eta}))|^2}+\frac{\xi^{2}}{v^{4}(1-\xi)^{2}}\right]\,,
\end{align}
\begin{align}
\sigma v_{\eta \eta \rightarrow \sigma \sigma}=\frac{m_{\eta}\sqrt{m^{2}_{\eta}-m^{2}_{\sigma}}((-7m^{4}_{\sigma}+2m^{2}_{\sigma}m^{2}_{\eta}+8m^{4}_{\eta})^{2}+(5m^{2}_{\sigma}-2m^{2}_{\eta})^{2}\Im(\Pi_{\sigma}(4m^{2}_{\eta}))^{2})}{4 \pi f^{4}(m^{2}_{\sigma}-2m^{2}_\eta )^{2} |m^{2}_{\sigma}-4m^{2}_{\eta}+i \Im(\Pi_{\sigma}(4m^{2}_{\eta}))|^2}\,,
\end{align}
\begin{align}
\sigma v_{\eta\eta\rightarrow \sigma h}=&\frac{\sqrt{m^{4}_{\sigma}+(-4m^{2}_{\eta}+m^{2}_{h})^{2}-2m^{2}_{\sigma}(4m^{2}_{\eta}+m^{2}_{h})}}{128 \pi f^{2} v^{2} m^{4}_{\eta} (1-\xi)}\left[\frac{4(m^{2}_{\sigma}+8m^{2}_{\eta}-m^{2}_{h})^{2}(m^{2}_{h}\xi -v^{2}\lambda_{h\eta}(1-\xi))^{2}}{(m^{2}_{\sigma}+m^{2}_{h}-4m^{2}_{\eta})^{2}}\right.\nonumber\\
&-\frac{4 (m^{2}_{\sigma}+8m^{2}_{\eta}-m^{2}_{h})(4\lambda_{h\eta}v^{2}+(m^{2}_{\sigma}-4m^{2}_{\eta}-m^{2}_{h}-4\lambda_{h\eta}v^{2})\xi)(m^{2}_{h}\xi-\lambda_{h\eta}v^{2}(1-\xi))}{m^{2}_{\sigma}-4m^{2}_{\eta}+m^{2}_{h}}\nonumber\\
&+(-4\lambda_{h\eta}v^{2}+(-m^{2}_{\sigma}+4m^{2}_{\eta}+m^{2}_{h}+\lambda_{h\eta}v^{2})\xi)^{2}+\frac{(m^{2}_{h}-4m^{2}_{\eta})^{2} (\lambda_{h\eta}v^{2}-(4m^{2}_{\eta}+\lambda_{h\eta}v^{2})\xi)^{2}}{ |m^{2}_{h}-4m^{2}_{\eta}+i \Im(\Pi_{h}(4m^{2}_{\eta}))|^2}\nonumber\\
&+\frac{4 (m^{2}_{\sigma}-4m^{2}_{\eta})(m^{2}_{h}-4m^{2}_{\eta})(8m^{2}_{\eta}+m^{2}_{\sigma}-m^{2}_{h})(-\lambda_{h\eta}v^{2}+(4m^{2}_{\eta}+\lambda_{h\eta})\xi)(-\lambda_{h\eta}v^{2}+(m^{2}_{h}+\lambda_{h\eta}v^{2})\xi)}{(4m^{2}_{\eta}-m^{2}_{\sigma}-m^{2}_{h})  |m^{2}_{h}-4m^{2}_{\eta}+i \Im(\Pi_{h}(4m^{2}_{\eta}))|^2}\nonumber\\
&\left.+\frac{2(-4\lambda_{h\eta}^{2}v^{4}+\lambda_{h\eta}v^{2}\xi(-m^{2}_{\sigma}+20m^{2}_{\eta}+m^{2}_{h}+8\lambda_{h\eta}v^{2})+\xi^{2}(4m^{2}_{\eta}+\lambda_{h\eta}v^{2})(m^{2}_{\sigma}-4m^{2}_{\eta}-m^{2}_{h}-4\lambda_{h\eta}v^2)}{(m^{2}_{\sigma}-4m^{2}_{\eta})^{-1}(m^{2}_{h}-4m^{2}_{\eta})^{-1} |m^{2}_{h}-4m^{2}_{\eta}+i \Im(\Pi_{h}(4m^{2}_{\eta}))|^2}\right]\,,
\end{align}
\begin{align}
\sigma v_{\eta\eta \rightarrow hh}=&\frac{m_{\eta} \sqrt{m^{2}_{\eta}-m^{2}_{h}}}{4 \pi m^{4}_{\eta} f^{8}}\left[\frac{64 v^{4}(m^{2}_{h}-f^{2}\lambda_{h\eta} (1-\xi))^{4}}{(2m^{2}_{\eta}-m^{2}_{h})^{2}(1-\xi)^{2}}+\frac{(4m^{2}_{\eta}-f^{2}\lambda_{h \eta}(1-\xi))^{2}(8(2m^{2}_{\eta}+m^{2}_{h})v^{2}-3f^{2}m^{2}_{h}(1-\xi))^{2}}{(1-\xi)^{2}|m^{2}_{h}-4m^{2}_{\eta}+i \Im(\Pi_{h}(4m^{2}_{\eta}))|^2}\right.\nonumber\\
&-\frac{16 (4m^{2}_{\eta}-m^{2}_{h})v^{2}(4m^{2}_{\eta}-f^{2}\lambda_{h\eta}(1-\xi))(m^{2}_{h}-f^{2}\lambda_{h\eta}(1-\xi))^{2}(8(2m^{2}_{\eta}+m^{2}_{h})v^{2}-3f^{2}m^{2}_{h}(1-\xi))}{(2m^{2}_{\eta}-m^{2}_{h})(1-\xi)^{2} |m^{2}_{h}-4m^{2}_{\eta}+i \Im(\Pi_{h}(4m^{2}_{\eta}))|^2}\nonumber\\
&+\frac{16 (m^{2}_{h}-f^{2}\lambda_{h\eta}(1-\xi))^{2}(8(m^{2}_{h}-2m^{2}_{\eta})v^{4}+f^{4}(-\lambda_{h\eta} v^{2}+\xi(m^{2}_{\eta}+2\lambda_{h\eta}v^{2})+\xi^{2}(m^{2}_{\eta}-\lambda_{h\eta}v^{2}))}{(2m^{2}_{\eta}-m^{2}_{h})(1-\xi)^{2}}\nonumber\\
&+\frac{2f^{4}(8(2m^{2}_{\eta}+m^{2}_{h})v^{2}-3f^{2}m^{2}_{h}(1-\xi))(8(-2m^{2}_{\eta}+m^{2}_{h})\xi^{2}-\lambda_{h\eta}v^{2}+\xi(m^{2}_{\eta}+2\lambda_{h\eta}v^{2})+\xi^{2}(m^{2}_{\eta}-\lambda_{h\eta}v^{2}))}{v^{2} (4m^{2}_{\eta}-m^{2}_{h})^{-1}(4m^{2}_{\eta}-f^{2}\lambda_{h\eta}(1-\xi))^{-1}|m^{2}_{h}-4m^{2}_{\eta}+i \Im(\Pi_{h}(4m^{2}_{\eta}))|^2 (1-\xi)^{2}}\nonumber\\
&+\frac{(8(2m^{2}_{\eta}-m^{2}_{h})v^{4}+f^{4}(\lambda_{h\eta}v^{2}-(m^{2}_{\eta}+2\lambda_{h\eta}v^{2})\xi+(-m^{2}_{\eta}+\lambda_{h\eta}v^{2})\xi^{2})^{2}}{v^{4}(1-\xi)^{4}}\nonumber\\
&-\frac{48 f^{2}v^{2}m^{2}_{\eta}(4m^{2}_{\eta}-m^{2}_{\sigma})(m^{2}_{h}-f^{2}\lambda_{h\eta}(1-\xi))^{2}(m^{2}_{h}+2m^{2}_{\eta})}{(2m^{2}_{\eta}-m^{2}_{h})|m^{2}_{\sigma}-4m^{2}_{\eta}+i \Im(\Pi_{\sigma}(4m^{2}_{\eta}))|^2 (1-\xi)}+\frac{9f^{4}m^{4}_{\eta}(2m^{2}_{\eta}+m^{2}_{h})^{2}}{|m^{2}_{\sigma}-4m^{2}_{\eta}+i \Im(\Pi_{\sigma}(4m^{2}_{\eta}))|^2}\nonumber\\
&+ \frac{6 f^{2}m^{2}_{\eta}((m^{2}_{\sigma}-4m^{2}_{\eta})(m^{2}_{h}-4m^{2}_{\eta})+\Im(\Pi_{h}(4m^{2}_{\eta}))\Im(\Pi(4m^{2}_{\eta}))) (8(2m^{2}_{\eta})v^{2}-3f^{2}m^{2}_{h}(1-\xi))(m^{2}_{h}+2m^{2}_{\eta})}{(4m^{2}_{\eta}-f^{2}\lambda_{h\eta}(1-\xi))^{-1} |m^{2}_{\sigma}-4m^{2}_{\eta}+i \Im(\Pi_{\sigma}(4m^{2}_{\eta}))|^2 |m^{2}_{h}-4m^{2}_{\eta}+i \Im(\Pi_{h}(4m^{2}_{\eta}))|^2 (1-\xi)}\nonumber\\
&\left.-\frac{6f^2 m^{2}_{\eta}(-m^{2}_{\sigma}+4m^{2}_{\eta})(2m^{2}_{\eta}+m^{2}_{h})(8(-2m^{2}_{\eta}+m^{2}_{h})v^{4}+f^{4}(-\lambda_{h\eta}v^{2}+\xi(m^{2}_{\eta}+2\lambda_{h\eta}v^{2})+\xi^{2}(m^{2}_{\eta}-\lambda_{h\eta}v^{2}))}{v^{2} (1-\xi)|m^{2}_{\sigma}-4m^{2}_{\eta}+i \Im(\Pi_{\sigma}(4m^{2}_{\eta}))|^2}\right]\,.
\end{align}
\end{widetext}

\end{document}